\documentclass{article}
\usepackage[utf8]{inputenc}
\usepackage[a4paper, total={6.5in, 9in}]{geometry}
\usepackage{amsmath,amsfonts,amssymb}
\usepackage{graphicx}
\usepackage{hyperref}
\usepackage{subcaption}
\usepackage{mathrsfs}
 \usepackage{textcomp} 
\usepackage[
    style=phys,
    sorting=none,
    backend=biber,
    giveninits=true
]{biblatex}

\usepackage{pgfplots}
\usepackage{float}
\usepackage{placeins}
\usepackage{multirow}
\usepackage{tikz}
\usetikzlibrary{trees,positioning}
\usepackage{forest}
\usepackage{booktabs}
\title{Transport Relaxation Mechanisms in Bilayer Graphene: Effects of Pauli Blocking}
\author{Amit Varshney, SSZ Ashraf \\Physics Department\\Faculty of Science\\
        Aligarh Muslim University, Aligarh-202002, India}
\date{\today}

\pgfplotsset{compat=1.18}

\begin{document}

\maketitle
\begin{abstract}
We present a unified analytical treatment of carrier transport
relaxation in bilayer graphene (BLG), including scattering by surface
roughness, charged and neutral impurities, acoustic and optical phonons,
and substrate polar phonons. The calculation is formulated within the
low-energy two-band approximation using Fermi's golden rule and the
semiclassical Boltzmann transport equation. Closed-form expressions for
the transport relaxation rates are derived and verified by numerical
evaluation of the corresponding scattering integrals. Particular
attention is given to Pauli blocking, whose influence becomes important
under degenerate carrier conditions owing to the approximately
parabolic low-energy dispersion and nearly constant density of states
of BLG. We find that Pauli blocking modifies the relaxation rates in a
mechanism- and energy-dependent manner for both suspended and
substrate-supported BLG. For the parameters considered, neutral-impurity
scattering provides the dominant contribution over a substantial
energy range, while acoustic-phonon and charged-impurity scattering
remain important competing mechanisms. Optical-phonon scattering is
suppressed below its emission threshold, whereas substrate polar
phonons provide an additional relaxation channel in supported BLG.
Comparison with monolayer graphene and a conventional two-dimensional
electron gas highlights the role of band dispersion, density of states,
screening, and chirality in determining the distinct scattering
behavior of BLG. These results provide analytical insight into the
relative importance of the principal momentum-relaxation mechanisms
and the role of Pauli blocking in BLG transport.
\end{abstract}

\section{Introduction}

Graphene is a two-dimensional form of carbon in which
\(\mathrm{sp}^2\)-bonded carbon atoms form a honeycomb lattice
\cite{castro2009electronic}. Although its unusual electronic properties
had been predicted theoretically, graphene was first isolated
experimentally in 2004 by Geim and Novoselov
\cite{znovoselov2004electric}. Subsequent experiments revealed an
unconventional integer quantum Hall effect, demonstrating that carriers
in monolayer graphene (MLG) behave as massless chiral quasiparticles
with a linear Dirac dispersion
\cite{novoselov2005two,zhang2005experimental}.

Bilayer graphene (BLG), most commonly realized in the Bernal (AB)
stacking configuration, has a markedly different low-energy electronic
structure. Interlayer coupling transforms the approximately linear
dispersion of MLG into an approximately parabolic low-energy spectrum,
so that the carriers behave as massive chiral quasiparticles. This
regime is described effectively by a two-band Hamiltonian
\cite{mccann2013electronic} and gives rise to an integer quantum Hall
effect distinct from that of MLG
\cite{mccann2006landau,novoselov2006unconventional}.

An important feature of BLG is its electrical tunability. Electrostatic
gating controls the carrier density, while a potential difference
between the two layers can open and tune a band gap
\cite{ohta2006controlling,mccann2006asymmetry,min2007ab}. Together with
its favorable electronic, thermal, mechanical, and optical properties
\cite{mccann2013electronic,balandin2011thermal,zhang2011mechanical,novoselov2012roadmap},
this tunability makes BLG attractive for nanoelectronic, optoelectronic,
thermoelectric, valleytronic, and high-frequency applications
\cite{oostinga2008gate,castro2007biased,xia2010graphene,yan2012dual,
wang2011enhanced,novoselov2012roadmap,abergel2010properties}.

For practical BLG devices, however, the electronic band structure alone
does not determine transport. Carrier momentum relaxation is governed
by a variety of disorder- and phonon-induced scattering mechanisms.
Important disorder mechanisms include neutral-impurity (NI),
surface-roughness (SR), and charged-impurity (CI) scattering, while
phonon-mediated processes include acoustic-phonon (AP), optical-phonon
(OP), and, in substrate-supported devices, surface-polar-phonon (SPP)
scattering \cite{peres2005,li2011electron,shah2024role,xiao2010charged}.
Substrate support also modifies the transport environment through
dielectric screening and introduces additional remote SPP scattering
channels \cite{fratini2008substrate,konar2010effect,li2010surface,
li2011electron}. In suspended graphene-based systems, flexural
out-of-plane (ZA) phonons can provide an additional low-energy
scattering channel, with their contribution depending on membrane
tension, anharmonic effects, and environmental constraints
\cite{Ochoa_BLG_Flexural_2011}. The present analysis, however, includes
only longitudinal acoustic  phonons  within the AP contribution and
does not explicitly treat ZA scattering.

The occupation of electronic final states provides another important
factor in transport scattering. Pauli blocking (PB), which follows from
the Pauli exclusion principle, reduces the phase space available for a
scattering transition when the final electronic state is occupied. This
effect is particularly relevant to inelastic processes, for which the
initial and final electronic states generally have different energies
and their occupations are governed by Fermi--Dirac statistics
\cite{bonaccorso2010graphene,arshia2021inelastic}. PB is expected to be
most pronounced in the degenerate regime and to weaken as increasing
temperature broadens the carrier distribution.

The analysis is restricted to momentum-relaxing disorder and
electron--phonon scattering. Normal electron--electron collisions are
not included because, in a translationally invariant system, they
conserve total electronic momentum and therefore do not directly
produce momentum relaxation. Such interactions can nevertheless be
important for carrier thermalization and hydrodynamic transport,
particularly near charge neutrality
\cite{wagner2020transport,fritz2024hydrodynamic}, but these effects lie
outside the scope of the present treatment. Likewise, flexural
out-of-plane (ZA) phonon scattering is not included; the AP
contribution considered here is restricted to longitudinal modes.

Carrier transport in BLG has been studied extensively for charged
impurities, short-range disorder, acoustic and optical phonons, and
substrate-related scattering mechanisms
\cite{das2010theory,li2011electron,xiao2010charged}. More recent work
has considered electron--phonon interactions, carrier cooling, and
hydrodynamic transport regimes
\cite{wagner2020transport,obaidurrahman2022cooling,fritz2024hydrodynamic,
claes2025phonon}. PB has also been investigated in detail for inelastic
transport in MLG \cite{arshia2021inelastic}. However, the combined
effects of electronic structure, momentum-dependent scattering
interactions, finite phonon energies, and final-state PB
are not commonly treated within a single analytical framework for
low-energy BLG transport.

In this work, we develop a unified analytical and numerical treatment
of the principal momentum-relaxing scattering mechanisms considered
here in low-energy BLG. NI, SR,
CI, AP, OP, and
SPP scattering are formulated within a common
Boltzmann transport framework. The treatment retains the momentum
dependence of finite-range scattering potentials and, for inelastic
processes, the finite phonon energies and associated emission and
absorption constraints. PB of the final electronic states
is incorporated explicitly, allowing its influence on the
energy-dependent transport relaxation rates to be quantified. The
resulting analytical and semi-analytical expressions are validated by
direct numerical evaluation of the corresponding scattering
integrals.

We further compare selected scattering rates in BLG with those of
MLG and a conventional two-dimensional electron
gas (2DEG). This comparison separates the effects of the scattering
interaction from those arising from the underlying band dispersion,
DOS, and chiral wave-function overlap. Finally, the
different scattering channels are evaluated over a common
carrier-energy range to identify their relative importance and the
regimes in which PB and finite-energy phase-space
restrictions substantially modify carrier relaxation. The resulting
framework provides a systematic basis for assessing how electronic
structure, scattering interaction, and final-state occupation
collectively govern transport relaxation in BLG.

\section{Formalism}

To investigate how different scattering mechanisms affect carrier transport in BLG, we use the semiclassical Boltzmann transport equation (BTE). Within the relaxation-time approximation, the BTE connects the nonequilibrium carrier distribution with the transport relaxation time \(\tau\). Its inverse, \(\tau^{-1}\), gives the corresponding transport scattering rate. Since not every scattering event contributes equally to electrical resistance, the transport rate includes an angular weighting factor that accounts for the amount of momentum lost in a scattering event. In particular, forward scattering produces little momentum relaxation, whereas backscattering is much more effective \cite{das2010theory,dassarma2011electronic}.

Following Refs.~\cite{hwang2008acoustic,arshia2021inelastic}, the
energy-dependent transport relaxation rate is written as
\begin{equation}
\frac{1}{\tau_k}
=
\frac{A}{(2\pi)^2}
\int k'\,dk'\,d\theta_{kk'}\,
(1-\cos\theta_{kk'})
\,t_{kk'}^{\mathrm{int}}
\frac{1-f(E_{k'})}
     {1-f(E_k)},
\label{eq:rel_rate}
\end{equation}
where \(A\) is the area of the BLG sheet, \(t_{kk'}^{\mathrm{int}}\)
denotes the interaction-induced transition probability for an electron to
scatter from the initial state \(\mathbf{k}\) to the final state
\(\mathbf{k}'\), and \(\theta_{kk'}\) is the scattering angle between the
two momentum states. The factor \(1-\cos\theta_{kk'}\) is the transport
factor and accounts for the efficiency with which a scattering event
relaxes the carrier momentum: it vanishes for forward scattering and
reaches its maximum for backscattering,
\(\theta_{kk'}=\pi\).

The ratio
\(
\frac{1-f(E_{k'})}{1-f(E_k)}
\)
is the PB factor and accounts for the occupation of the
final electronic state in the scattering process
\cite{arshia2021inelastic}. For strictly elastic scattering, \(E_{k'}=E_k\), and hence the PB
factor reduces identically to unity. Its nontrivial contribution in
the present formulation therefore arises in inelastic phonon-mediated
processes. The electronic occupation is described by the
Fermi--Dirac distribution,
\(
f(E)
=
\frac{1}{1+\exp[\beta(E-\mu)]},
\qquad
\beta=\frac{1}{k_BT},
\)
where \(\mu\) is the chemical potential, \(k_B\) is the Boltzmann constant,
and \(T\) is the absolute temperature.

Within the low-energy two-band approximation, BLG is described by an
approximately parabolic dispersion,
\(
E_k=\frac{\hbar^2k^2}{2m^*},
\)
where \(m^*\) is the effective electron mass
\cite{mccann2013electronic,das2010theory}. The corresponding low-energy
spinor wavefunction may be written as
\(
\psi_{\mathbf{k}}(\mathbf{r})
=
\frac{1}{\sqrt{2A}}
\begin{pmatrix}
e^{i\theta_k}\\
e^{-i\theta_k}
\end{pmatrix}
e^{i\mathbf{k}\cdot\mathbf{r}},
\)where \(\mathbf{r}\) is the position and \(\theta_k\) is the polar angle of the electron wave vector
\(\mathbf{k}\). The winding-two chiral structure of the BLG spinor gives
rise to the overlap factor
\(u_{kk'}
=
\left|
\langle u_{\mathbf{k}'}|u_{\mathbf{k}}\rangle
\right|^2
=
\frac{1+\cos(2\theta_{kk'})}{2},
\)
which determines the angular dependence of scattering in the low-energy BLG model
\cite{mccann2013electronic}.

For a scattering process from \(\mathbf{k}\) to \(\mathbf{k}'\), the
transferred momentum is
\(
q
=
|\mathbf{k}'-\mathbf{k}|
=
\sqrt{k^2+k'^2-2kk'\cos\theta_{kk'}}.
\)
For elastic scattering, \(E_{k'}=E_k\), and hence \(k'=k\), so that
\(
q
=
2k\sin\left(\frac{\theta_{kk'}}{2}\right).
\)
For inelastic scattering, \(k\) and \(k'\) are generally different and
are related through the appropriate energy-conservation condition.

The transition probability is evaluated using Fermi's golden rule,
\begin{equation}
t_{kk'}^{\mathrm{int}}
=
\frac{2\pi}{\hbar}
\left|
\langle k'|H_{\mathrm{int}}|k\rangle
\right|^2
\delta(E_f-E_i),
\label{eq:fermi}
\end{equation}
where \(E_i\) and \(E_f\) denote the total initial and final energies of
the electron--scattering-system configuration, respectively
\cite{das2010theory}. For elastic impurity scattering,
\(E_f-E_i=E_{k'}-E_k\), whereas phonon-mediated scattering additionally
involves the absorption or emission of a phonon with energy
\(\hbar\omega_q\).

For the scattering mechanisms considered here, the squared interaction
matrix element can be separated into a mechanism-dependent coupling and
the BLG chiral overlap factor,
\begin{equation}
|v_{kk'}|^2
=
|M(q)|^2u_{kk'},
\label{eq:matrix_element_general}
\end{equation}
where \(M(q)\) contains the mechanism-specific coupling.

For phonon-mediated processes, the electron--phonon interaction can be
written in the generic second-quantized form
\(
H_{e\text{-}ph}
=
\sum_{\mathbf{k},\mathbf{q}}
v_{\mathbf{k}\mathbf{q}}\,
c_{\mathbf{k}+\mathbf{q}}^{\dagger}
c_{\mathbf{k}}
\left(
b_{\mathbf{q}}+
b_{-\mathbf{q}}^{\dagger}
\right),
\)
where \(c_{\mathbf{k}}^\dagger\) and \(c_{\mathbf{k}}\) are electron
creation and annihilation operators, while
\(b_{\mathbf{q}}^\dagger\) and \(b_{\mathbf{q}}\) are the corresponding
phonon operators \cite{li2011electron,arshia2021inelastic}.

For elastic disorder-induced scattering, the transition rate can be
written in the general form
\begin{equation}
t_{kk'}^{\mathrm{dis}}
=
\frac{2\pi}{\hbar}
\mathcal{W}_{\mathrm{dis}}(q)
u_{kk'}
\delta(E_{k'}-E_k),
\label{eq:disorder_transition_general}
\end{equation}
where \(\mathcal{W}_{\mathrm{dis}}(q)\) denotes the
mechanism-dependent disorder coupling function. It incorporates the
strength and spatial correlations of the disorder and, where appropriate,
the effects of dielectric screening. For a specific disorder mechanism,
\(\mathcal{W}_{\mathrm{dis}}(q)\) is determined by the corresponding
disorder potential or disorder correlation function
\cite{das2010theory,xiao2010charged}. The specific forms of
\(\mathcal{W}_{\mathrm{dis}}(q)\) for NI,
SR, and CI scattering are derived separately
in the corresponding sections.

For phonon-mediated scattering, the transition probability obtained from
Fermi's golden rule is
\begin{equation}
t_{kk'}^{\mathrm{ph}}
=
\frac{2\pi}{\hbar}
|v_{kq}|^2
\left[
n_{\rm ph}\,
\delta\!\left(
E_k+\hbar\omega_q-E_{k'}
\right)
+
(n_{\rm ph}+1)\,
\delta\!\left(
E_k-\hbar\omega_q-E_{k'}
\right)
\right],
\label{eq:phonon_transition}
\end{equation}
where \(n_{\rm ph}\) is the equilibrium phonon occupation number,
\(
n_{\rm ph}
=
\frac{1}
{\exp(\beta\hbar\omega_q)-1},
\)
which follows the Bose--Einstein distribution
\cite{arshia2021inelastic}. The first term in
Eq.~(\ref{eq:phonon_transition}) describes phonon absorption, for which
the electron gains energy \(\hbar\omega_q\), whereas the second term
describes phonon emission, for which the electron loses the same amount
of energy. The Dirac delta functions enforce energy conservation for the
respective processes. For OPs with approximately fixed energy
$\hbar\omega_o$, the emission term is subject to a finite-energy
threshold, whereas AP scattering involves much smaller
phonon energies and approaches the quasi-elastic limit when
$\hbar\omega_{\rm AP}\ll k_BT$.

The Coulomb interaction associated with charged scattering centers is
modified by dielectric screening. Within the random-phase
approximation, the static dielectric function can be written as
\begin{equation}
\epsilon_q
=
1+
\frac{2\pi e_c^2}
{\epsilon q}
\Pi(q),
\label{eq:dielectric}
\end{equation}
where \(\epsilon\) denotes the effective background dielectric constant
of the BLG--substrate environment and \(\Pi(q)\) is the static
polarization function of BLG
\cite{das2010theory,hwang2009screening}.
In the long-wavelength Thomas--Fermi approximation, the static polarization
function is replaced by its long-wavelength limit,
\(\Pi(q)\simeq\Pi(0)\simeq D(E_F),\)
where \(D(E_F)\) is the DOS at the Fermi energy. This approximation is appropriate in the degenerate, long-wavelength regime. The resulting screened dielectric function enters the matrix elements of scattering mechanisms whose interaction potentials are affected by the dielectric response of the BLG environment, most notably CI scattering and, where applicable, SR scattering.

The scattering mechanisms considered in this work are intended to capture the principal disorder- and phonon-induced processes responsible for carrier momentum relaxation over the energy and temperature ranges studied. The disorder-related mechanisms are NI, SR, and CI scattering. These represent, respectively, short-range neutral disorder, potential fluctuations associated with the interface, and long-range Coulomb disorder. The phonon-related mechanisms are AP, OP, and SPP scatterings, representing intrinsic lattice vibrations and polar phonon modes associated with the substrate. Together, these mechanisms constitute the main elastic and inelastic scattering channels included in the present BLG transport model. Other possible mechanisms are not considered because their treatment would require additional material-specific parameters or would extend beyond the scope of the present analysis.

The corresponding squared interaction coupling functions are introduced below:
\begin{equation}
|v_{kk'}|^2
=
\begin{cases}
\displaystyle
\left(2\pi a^2 V_0\right)^2
e^{-q^2a^2}
u_{kk'},
&
\text{(NI)\;\cite{peres2005}.}
\\[3ex]

\displaystyle
\frac{
\pi d^2 \Delta^2
\left(
\frac{2\pi e_c^2 n_{\rm int}}{\epsilon_s}
\right)^2
\exp\!\left(-\frac{q^2 d^2}{4}\right)
}{
\epsilon_q^2
}
u_{kk'},
&
\text{(SR)\;\cite{shah2024role}.}
\\[3ex]

\displaystyle
\frac{
\left(
\frac{2\pi e_c^2}{\epsilon_s qA}
\right)^2
\left[
e^{-2qb}
+
e^{-2q(b+l)}
\right]
}{
\epsilon_q^2
}
u_{kk'},
&
\text{(CI)\;\cite{das2010theory,xiao2010charged}.}
\\[3ex]

\displaystyle
\frac{
V_a^2\hbar q
}{
2A\rho_b v_{lb}
}
u_{kk'},
&
\text{(AP)\;\cite{li2011electron,arshia2021inelastic}.}
\\[3ex]

\displaystyle
\frac{
V_o^2\hbar
}{
A\rho_b\omega_o
}
u_{kk'},
&
\text{(OP)\;\cite{li2011electron,arshia2021inelastic}.}
\\[3ex]

\displaystyle
\frac{
V_s e_c^2
\left[
e^{-2qb}
+
e^{-2q(b+l)}
\right]
}{
Aq
}
u_{kk'},
&
\text{(SPP)\;\cite{li2010surface,li2011electron}.}
\end{cases}
\label{eq:coupling_functions}
\end{equation}

Here, \(v_{kk'}\) denotes the effective interaction matrix element for the
corresponding scattering mechanism, including the BLG chiral overlap
factor \(u_{kk'}\). The mechanism-dependent part of the matrix element
contains the relevant interaction strength, momentum dependence, disorder
correlations, phonon normalization, and, where appropriate, dielectric
screening. The explicit forms also retain the normalization-area factors
associated with the particular interaction Hamiltonians. These factors
must be treated consistently when the matrix elements are inserted into
the Fermi-golden-rule transition rates. In particular, the final
scattering rates must have dimensions of inverse time and must be
independent of the arbitrary normalization area \(A\).

For the inelastic phonon mechanisms, the appropriate phonon energy
$\hbar\omega_q$ enters the transition probability through
Eq.~(\ref{eq:phonon_transition}), whereas for elastic mechanisms the
energy-conservation condition reduces to \(E_{k'}=E_k\)
\cite{arshia2021inelastic}.

The quantities \(V_a\) and \(V_o\) denote the deformation-potential coupling
strengths for APs and OPs, respectively, whereas \(V_s\)
characterizes the coupling to substrate SPPs
\cite{li2010surface}. The SPP coupling parameter is given
by
\begin{equation}
V_s
=
\frac{\hbar\omega_{sp}}{2}
\left(
\frac{1}{\epsilon_h+\epsilon_0}
-
\frac{1}{\epsilon_l+\epsilon_0}
\right),
\label{eq:vs}
\end{equation}
where \(\epsilon_h\) and \(\epsilon_l\) denote the high- and low-frequency
dielectric parameters of the substrate, respectively, and
\(\epsilon_0\) is the permittivity of free space.

The transport relaxation time for each scattering mechanism is evaluated
independently using Eqs.~(\ref{eq:rel_rate})--(\ref{eq:coupling_functions}),
together with the corresponding dielectric screening function. When the
different scattering mechanisms can be treated as independent, their
transport scattering rates are additive according to Matthiessen's rule,
\(\frac{1}{\tau_{\mathrm{tot}}(E)}
=
\sum_i \frac{1}{\tau_i(E)},
\)
where the sum extends over the scattering mechanisms considered here:
NI, SR, CI, AP,
OP, and SPP scattering.

The summary of the principal carrier-scattering mechanisms
considered in BLG with the
underlying interaction, characteristic energy scale and scattering
process, relative importance, and conditions under which each
mechanism can become significant has been tabulated in  Table~\ref{tab:scattering}. 

\begin{table*}[h!]
\centering
\renewcommand{\arraystretch}{1.25}
\small

\begin{tabular}{|p{1.8cm}|p{2.0cm}|p{2.5cm}|p{1.8cm}|p{5.0cm}|}
\hline
\textbf{Scattering mechanism} &
\textbf{Interaction} &
\textbf{Energy scale / process} &
\textbf{Relative importance} &
\textbf{Conditions / representative references} \\
\hline

Neutral impurity (NI) &
Short-range disorder &
Elastic; no characteristic phonon energy &
Low--moderate &
Important in the presence of short-range neutral defects or
adsorbates; sensitive to impurity density and correlation length.
\cite{lewenkopf2013computational,ortmann2015graphene,ferreira2011unified} \\
\hline

Surface roughness (SR) &
Interface disorder &
Elastic; no characteristic phonon energy &
Low--moderate &
Relevant for rough or imperfect dielectric interfaces; depends
strongly on roughness amplitude and correlation length.
\cite{shishir2009room,shah2024role} \\
\hline

Charged impurity (CI) &
Screened Coulomb interaction &
Elastic; no characteristic phonon energy &
Moderate--high &
Often important at low carrier density; strongly dependent on charged
impurity density, dielectric environment, screening, and impurity
distance.
\cite{hwang2009screening,adam2007self,dassarma2011electronic} \\
\hline

Acoustic phonon (AP) &
Deformation potential &
Quasi-elastic; $\hbar\omega_q\ll k_BT$ &
Moderate &
Becomes increasingly important with temperature and can dominate
intrinsic transport in relatively clean samples.
\cite{kubakaddi2009interaction,hwang2008acoustic} \\
\hline

Intrinsic optical phonon (OP) &
Deformation potential &
Inelastic; $\hbar\omega_o$ &
Low below threshold; high at elevated carrier energies &
Important for hot carriers and high electric fields; emission requires
$E_k\geq\hbar\omega_o$. Below the emission threshold, the OP
contribution is strongly suppressed by restricted inelastic phase
space.
\cite{borysenko2011electron,park2014electron,tse2009energy,borysenko2010first} \\
\hline

Surface polar phonon (SPP) &
Fr\"ohlich interaction &
Inelastic; $\hbar\omega_{sp}$ &
Moderate--high &
Important in graphene supported on polar or high-$\kappa$ dielectric
substrates; strongly dependent on substrate phonon energy and
graphene--substrate separation.
\cite{fratini2008substrate,konar2010effect,ong2012theory} \\
\hline

\textbf{Total} &
Combined scattering &
Combined elastic and inelastic processes &
System dependent &
Total transport rate obtained from Matthiessen's rule,
\(
\tau_{\mathrm{tot}}^{-1}=\sum_i\tau_i^{-1}.
\)
The dominant mechanism depends on carrier density, temperature,
substrate, disorder, and carrier energy.
\cite{dassarma2011electronic} \\
\hline

\end{tabular}

\caption{Summary of the principal carrier-scattering mechanisms
considered in bilayer graphene (BLG). The table specifies the
underlying interaction, characteristic energy scale and scattering
process, relative importance, and conditions under which each
mechanism can become significant. The relative importance is
qualitative and depends on carrier density, temperature, substrate,
disorder strength, carrier energy, and the parameters characterizing
the electron--phonon or impurity coupling. The corresponding coupling
matrix elements are given in Eq.~(\ref{eq:coupling_functions}). The
total transport scattering time is obtained from Matthiessen's rule,
\(
\tau_{\mathrm{tot}}^{-1}=\sum_i\tau_i^{-1}.
\)}
\label{tab:scattering}
\end{table*}

\section{Neutral Impurity Scattering}
In this section, we consider elastic scattering by weak finite-range
NIs within the framework of the first Born approximation.
Such a model can represent weak neutral disorder associated with
substitutional defects, weak neutral adsorbates, impurity clusters, and
smooth spatially correlated disorder.
\cite{peres2010transport,mccann2013electronic}.
The NI matrix element in
Eq.~(\ref{eq:disorder_transition_general}), obtained by substituting the
NI coupling function from Eq.~(\ref{eq:coupling_functions}),
can be evaluated explicitly for a finite-range Gaussian impurity
potential, yielding an analytical expression for the corresponding
transport relaxation time. The impurity
potential is modeled as
\(V(r)
=
V_0\exp\left(-\frac{r^2}{2a^2}\right),
\) where \(V_0\) denotes the amplitude of the impurity potential and \(a\)
characterizes its spatial extent. Finite-range Gaussian potentials are widely used as model disorder
potentials to describe spatially correlated, smooth neutral disorder in
graphene-based transport calculations. This model is appropriate for
weak disorder, for which multiple-scattering effects are negligible and
the first Born approximation is applicable. It is not intended to describe resonant
scatterers, such as vacancies or hydrogen adatoms, whose strong localized
potentials require a non-perturbative $T$-matrix treatment
\cite{ferreira2011unified}.

The two-dimensional Fourier transform of the Gaussian impurity potential
is
\(V(q)
=
2\pi a^2V_0
\exp\left(-\frac{q^2a^2}{2}\right),
\)
and hence
\(
|V(q)|^2
=
(2\pi a^2V_0)^2
\exp(-q^2a^2).
\) The effective squared coupling function for NI is written in Eq.~(\ref{eq:coupling_functions}) \(|v_{kk'}|^2
=\left(2\pi a^2 V_0\right)^2
e^{-q^2a^2}
u_{kk'},   
\) 
Using the NI transition probability from
Eq.~(\ref{eq:coupling_functions}) in the general transport relaxation-rate
expression of Eq.~(\ref{eq:rel_rate}) and, for the elastic NI contribution in the present
transport formulation, the PB factor reduces to unity. The transport
relaxation rate at the Fermi energy is therefore
\begin{equation}
\frac{1}{\tau_{NI}(E_F)}
=
\frac{2\pi n_{\rm NI}g_s g_v}{\hbar}
\int
\frac{d^2k'}{(2\pi)^2}
|V(q)|^2
u_{kk'}
\left(1-\cos\theta_{kk'}\right)
\delta(E_k-E_{k'}),
\label{eq:tau_general}
\end{equation}
where \(n_{\rm NI}\) is the NI density.

For the transport rate evaluated at the Fermi energy, elastic energy
conservation gives \(k=k'=k_F\), and therefore
\(
q
=
2k_F\sin\left(\frac{\theta_{kk'}}{2}\right).
\)
Consequently,
\(
|V(q)|^2
=
(2\pi a^2V_0)^2
\exp\left[
-4k_F^2a^2
\sin^2\left(\frac{\theta_{kk'}}{2}\right)
\right].
\)

Using the parabolic BLG dispersion and the chiral overlap factor \cite{mccann2013electronic}
\(
u_{kk'}
=
\frac{1+\cos(2\theta_{kk'})}{2}
=
\cos^2\theta_{kk'},
\)
the scattering rate becomes
\begin{equation}
\frac{1}{\tau_{NI}}
=
\frac{2\pi n_{\rm NI}}{\hbar}
(2\pi a^2V_0)^2
D(E_F)
\frac{1}{2\pi}
\int_0^{2\pi}
e^{-4k_F^2a^2\sin^2(\theta/2)}
\cos^2\theta
(1-\cos\theta)
\,d\theta,
\label{eq:tau_angle}
\end{equation}
where
\(D(E_F)
=
\frac{g_sg_vm^*}{2\pi\hbar^2}
=
\frac{2m^*}{\pi\hbar^2},
\)
is the total DOS including spin and valley degeneracies,
with \(g_s=g_v=2\).

Introducing
\(
x=2k_F^2a^2,
\)
and using
\(
e^{-4k_F^2a^2\sin^2(\theta_{kk'}/2)}
=
e^{-x}e^{x\cos\theta_{kk'}},
\)
together with
\(
\cos^2\theta_{kk'}(1-\cos\theta_{kk'})
=
\frac12-\frac34\cos\theta_{kk'}
+\frac12\cos2\theta_{kk'}
-\frac14\cos3\theta_{kk'},
\)
and
\(
\int_0^{2\pi}
e^{x\cos\theta_{kk'}}\cos(n\theta_{kk'})\,d\theta_{kk'}
=
2\pi I_n(x),
\)
where \(I_n(x)\) is the modified Bessel function of the first kind of
order \(n\), yields
\begin{equation}
\frac{1}{\tau_{NI}}
=
\frac{2\pi n_{\rm NI}}{\hbar}
(2\pi a^2V_0)^2
D(E_F)
e^{-x}
\left[
\frac12I_0(x)
-\frac34I_1(x)
+\frac12I_2(x)
-\frac14I_3(x)
\right].
\label{eq:tau_final}
\end{equation}

In the short-range limit, the Gaussian potential approaches a contact
potential provided the integrated potential strength
\(
U_0=2\pi a^2V_0
\)
is held constant as \(a\rightarrow0\). Since
\(
I_0(0)=1,
\qquad
I_n(0)=0,\quad n\geq1,
\)
Eq.~(\ref{eq:tau_final}) reduces to
\begin{equation}
\frac{1}{\tau_{NI}^{(0)}}
=
\frac{\pi n_{\rm NI}}{\hbar}
U_0^2D(E_F)=
\frac{2n_{\rm NI}m^*U_0^2}{\hbar^3}.,
\label{eq:tau_short_range}
\end{equation}

Thus, within the low-energy two-band approximation, the short-range
NI contribution gives an energy-independent transport
scattering rate, consistent with the established behavior of
short-range disorder in BLG \cite{mccann2013electronic,ferreira2011unified}.

\section{Surface Roughness Scattering}

In practical substrate-supported BLG devices, the graphene--dielectric
interface is not atomically flat but exhibits random height fluctuations.
These fluctuations modify the local electrostatic potential experienced
by charge carriers and can therefore contribute to SR
scattering 
\cite{sakaki1987interface,shishir2009room,shah2024role}.
The corresponding squared coupling function in
Eq.~(\ref{eq:coupling_functions}) is
\(
|v_{\rm kk'}^{SR}|^2
=
\frac{
\pi d^{2}\Delta^{2}
\left(
\dfrac{2\pi e_c^{2}n_{\rm int}}{\epsilon_s}
\right)^{2}
\exp\left(-\dfrac{q^{2}d^{2}}{4}\right)
}{
\epsilon_q^{2}
}u_{kk'},
\)
as adopted in SR models for graphene-based structures
\cite{shishir2009room,shah2024role}.
For elastic SR, \(k'=k=k_F\). It is convenient to transform the angular integral into an integral
over the momentum transfer
\(q=|\mathbf{k}'-\mathbf{k}|
=
2k_F\sin\left(\frac{\theta_{kk'}}{2}\right).
\)
Consequently,
\(
\begin{aligned}
(1-\cos\theta_{kk'})u_{kk'}d\theta_{kk'}
=
\frac{q^2dq}
{2k_F^3\sqrt{1-q^2/(4k_F^2)}}
\left[
1-\frac{q^2}{k_F^2}
\left(
1-\frac{q^2}{4k_F^2}
\right)
\right].
\end{aligned}
\)
The two angular branches corresponding to
\(\theta_{kk'}\) and \(2\pi-\theta_{kk'}\) give identical contributions.
Thus, the full angular integral over \(0\leq\theta_{kk'}\leq2\pi\) can be
reduced to an integral over \(0\leq q\leq2k_F\), with the corresponding symmetry factor included in the prefactor.

The resulting general screened SR rate is
\begin{equation}
\begin{aligned}
\frac{1}{\tau_{\rm SR}}
={}&
\pi d^{2}\Delta^{2}
\left(
\frac{2\pi e_c^{2}n_{\rm int}}{\epsilon_s}
\right)^{2}
\frac{D({E_F)}}
{2\hbar k_F^{3}}
\
&\times
\int_{0}^{2k_F}
\frac{
\exp\left(-\dfrac{q^{2}d^{2}}{4}\right)
}{
\epsilon_q^{2}
}
\frac{
q^{2}
}{
\sqrt{1-\dfrac{q^{2}}{4k_F^{2}}}
}
\left[
1-
\frac{q^{2}}{k_F^{2}}
\left(
1-\frac{q^{2}}{4k_F^{2}}
\right)
\right]
dq .
\end{aligned}
\label{eq:SRSR}
\end{equation}
For the screened case,  and after evaluating the
momentum integral analytically, Eq.~\eqref{eq:SRSR}
reduces to

\begin{equation}
    \frac{1}{\tau_{\rm SR}}
=\frac{\pi ^2 \Delta^2 n_s^2 \hbar }{m^*d^2  D(E_F) E_k}e^{-K_s} \left(8(2K_s^2+2K_s+3)Bl[0,K_s]-(4K_s^2 + 8)(2K_s + 3) {}_0F_1\left[2,\frac{K_s^2}{4}\right]\right)
    \label{eq:ASRSR} 
\end{equation}
where \(K_s=\frac{m^*d^2E_k}{\hbar^2}, \ Bl[i,x]\) denote the \(i^{\text{th}}\)-order Bessel function and \({}_0 F_1[a,x]=\sum_0^\infty \frac{1}{(a)_k} \frac{x^k}{k!}\) is regularised Hypergeometric function, with argument \(x\).
Equation~\eqref{eq:ASRSR} con
stitutes the general screened SR scattering rate
within the present model. The exponential factor arises from the assumed Gaussian height--height
correlation of the SR and therefore determines the momentum dependence of the roughness spectrum.

To obtain an analytical expression, we first consider the unscreened
limit, \(\epsilon_q=1\). Using
\(
q=2k_F\sin\left(\frac{\theta_{kk'}}{2}\right),
\)
the Gaussian factor becomes
\(
\exp\left[
-k_F^2d^2\sin^2\left(\frac{\theta_{kk'}}{2}\right)
\right]
=
e^{-x_{\rm SR}}
e^{x_{\rm SR}\cos\theta_{kk'}},
\)
where
\(
x_{\rm SR}=\frac{k_F^2d^2}{2}.
\)

Thus, the unscreened SR scattering rate becomes
\begin{equation}
\frac{1}{\tau_{\rm SR}^{(0)}}
=
\frac{\pi d^{2}\Delta^{2}
D{(E_F)}}
{4\hbar}
\left(
\frac{2\pi e_c^{2}n_{\rm int}}{\epsilon_s}
\right)^2
\int_0^{2\pi}
e^{-x_{\rm SR}}
e^{x_{\rm SR}\cos\theta_{kk'}}
\cos^2\theta_{kk'}(1-\cos\theta_{kk'})
\,d\theta_{kk'}.
\label{eq:SRS_angular}
\end{equation}

Using
\(
\int_0^{2\pi}
e^{x\cos\theta}\cos(n\theta)\,d\theta
=
2\pi I_n(x),
\)
where \(I_n(x)\) is the modified Bessel function of the first kind of
order \(n\), Eq.~(\ref{eq:SRS_angular}) yields
\begin{equation}
\frac{1}{\tau_{\rm SR}^{(0)}}
=
\frac{\pi^{2}d^{2}\Delta^{2}}
{2\hbar}
\left(
\frac{2\pi e_c^{2}n_{\rm int}}{\epsilon_s}
\right)^2
D_{(E_F)}
e^{-x_{\rm SR}}
\left[
\frac12I_0(x_{\rm SR})
-\frac34I_1(x_{\rm SR})
+\frac12I_2(x_{\rm SR})
-\frac14I_3(x_{\rm SR})
\right].
\label{eq:SRS_analytical}
\end{equation}

In the short-correlation-length limit,
\(x_{\rm SR}\rightarrow0\), using
\(I_0(0)=1,\qquad I_n(0)=0,\quad n\geq1,\)
Eq.~(\ref{eq:SRS_analytical}) reduces to
\begin{equation}
\frac{1}{\tau_{\rm SR}^{(0)}}
=
\frac{\pi^{2}d^{2}\Delta^{2}}
{4\hbar}
\left(
\frac{2\pi e_c^{2}n_{\rm int}}{\epsilon_s}
\right)^2
D{(E_F)}.
\label{eq:SRS_short_range}
\end{equation}

Thus, in the absence of dielectric screening, the SR scattering rate admits a
closed-form analytical expression in terms of modified Bessel
functions. Its functional form is analogous to that obtained for
finite-range NI scattering because both mechanisms
contain Gaussian momentum-dependent factors and the same BLG chiral
overlap factor. Their physical origins and coupling prefactors,
however, are distinct: the Gaussian factor in NI
scattering originates from the finite spatial range of an individual
impurity potential, whereas in SR scattering it arises from the statistical
height--height correlation function of the graphene--substrate
interface.

For the full momentum-dependent screening function \(\epsilon_q\), the
factor \(\epsilon_q^{-2}\) prevents the angular integral from reducing
to the same finite combination of modified Bessel functions. The fully
screened SR scattering rate is therefore evaluated numerically from
Eq.~(\ref{eq:SRSR}).

\section{Charged Impurity Scattering}

CI scattering is an important mechanism limiting carrier
mobility in BLG, particularly in devices supported on dielectric
substrates such as SiO$_2$.
\cite{das2010theory,xiao2010charged} CIs
arising from substrate defects, adsorbed ions, or fabrication-induced
contaminants generate long-range Coulomb potentials that scatter charge
carriers and degrade electronic transport. Screened CI scattering has been extensively studied in
graphene transport, particularly in MLG.
\cite{adam2009theory,hwang2007carrier}
For BLG, the corresponding transport rate incorporates the parabolic
low-energy dispersion, the BLG chiral overlap factor, and dielectric
screening.\cite{das2010theory}

For CI scattering, the mechanism-specific matrix element
is obtained from the screened Coulomb interaction together with the
distance-dependent form factor associated with the impurity location.\cite{das2010theory, xiao2010charged}
Using this matrix element in the general transport-rate expression Eq. (\ref {eq:rel_rate}) gives
\begin{equation}
\frac{1}{\tau_{\rm CI}}
=
\frac{2m^*n_{\rm CI}g_s g_v}{\pi\hbar^3}
\int_0^{2\pi}
\left(
\frac{2\pi e_c^2}{\epsilon_s q}
\right)^2
\left(
1+\frac{4m^*e_c^2}
{\epsilon_s\hbar^2q}
\right)^{-2}
\left[
e^{-2qb}+e^{-2q(b+l)}
\right]
\frac{1+\cos 2\theta_{kk'}}{2}
\left(1-\cos\theta_{kk'}\right)
\,d\theta_{kk'},
\label{eq:CI_integral}
\end{equation}
where \(n_{\rm CI}\) is the CI density, \(q=2k\sin\left(\frac{\theta_{kk'}}{2}\right)\), \(E_{k'}=E_k\), and the PB ratio is unity 
for elastic scattering. The Thomas--Fermi screening wave vector is \(q_{\rm TF}
=
\frac{4m^*e_c^2}{\epsilon_s\hbar^2}.
\)

Here the two exponential terms represent the contributions associated
with CIs located at distances \(b\) and \(b+l\) from the
reference BLG layer. The expression assumes statistically independent
contributions from the two impurity populations. If coherent
interference between the corresponding scattering amplitudes is
included, the distance-dependent form factor must instead be modified
to contain the appropriate cross term.

The angular integral in Eq.~(\ref{eq:CI_integral}) is difficult to
evaluate analytically in its general form. But under strong screening, \(\alpha=\frac{q_{\rm TF}}{2k}\gg1\), the resulting integral Eq.~(\ref{eq:CI_integral})can be evaluated analytically , giving

\begin{multline}
\frac{1 }{\tau_{CI}}= \mathscr{A(\gamma)} + \mathscr{A}(\Gamma) \\ where \\
\mathscr{A}(\gamma)=-\frac{1}{48 \gamma^4 m^*} \pi \ n_{CI} \hbar \Big(-32\gamma^3 (3+4\gamma^2) + 9\pi\gamma(5+4\gamma^2) \ (Sl[1,4\gamma]\ -\ Bl[1,4\gamma] )+ 3\pi(15+25\gamma^2 +16\gamma^4)\ (Bl[2,4\gamma] \\ - \ Sl[2,4\gamma])\Big)
\label{eq:CISE}
\end{multline}
Where
\(\gamma=(d/\hbar)\sqrt{2m^*E_k}, \qquad
\Gamma=((c+d)/\hbar)\sqrt{2m^*E_k}\),
where \(Bl[i,x]\) and \(Sl[i,x]\) denote the \(i^{\text{th}}\)-order Bessel and Struve functions, respectively, with argument \(x\).
Equation~(\ref{eq:CISE}) provides the analytical CI
relaxation rate. For the parabolic BLG dispersion,
\(k=\frac{\sqrt{2m^*E_k}}{\hbar},\)
so that \(\alpha\propto E_k^{-1/2}\).

If screening is removed then the Eq.~(\ref{eq:CI_integral}) yields the following analytical result
\begin{multline}
    \frac{\pi ^2 e_c^4 n_{CI} 
    }{16 E_k \epsilon_s^2 \hbar }\left(\mathscr{B(\gamma)+B}(\Gamma)\right) \\ where \\
    \mathscr{B(\gamma)}=\frac{1}{\gamma^3}\left(-24\gamma \ Sl[2,4\gamma] + (64\gamma^3 + 24\gamma)Bl[0,4\gamma]-(32\gamma^2+12)Bl[1,4\gamma]+32\gamma^2 \ Sl[1,4\gamma] - 32\gamma^3 \ Sl[0,4\gamma]\right)
    \label{eq:CIUSE}
\end{multline}
A particularly simple analytical result follows if \(b=l=0\) in the exponential factors in Eq.~(\ref{eq:CI_integral}) and in the strong-screening
regime, where 
\(q_{\rm TF}\gg q.\)
In this limit,
\(\left(1+\frac{q_{\rm TF}}{q}\right)^{-2}
\simeq
\frac{q^2}{q_{\rm TF}^2},
\)
and hence
\begin{equation} 
\frac{1}{\tau_{\rm CI}^{(0)}}
\simeq
\frac{4\pi m^*n_{\rm CI}e_c^4}
{\epsilon_s^2\hbar^3q_{\rm TF}^2}
\int_0^{2\pi}
\cos^2\theta_{kk'}
(1-\cos\theta_{kk'})
\,d\theta_{kk'} \nonumber
=
\frac{4\pi m^*n_{\rm CI}e_c^4}
{\epsilon_s^2\hbar^3q_{\rm TF}^2}\,\pi
\nonumber
=
\frac{\pi^2}{4}
\frac{\hbar n_{\rm CI}}{m^*}.
\end{equation}

Thus, within the strong-screening and zero-separation approximations, the CI scattering rate becomes independent of the carrier wave vector \(k\) to leading order. The explicit dependence on the dielectric constant
and Coulomb interaction strength cancels through the Thomas--Fermi
screening wave vector, leaving a rate proportional to the CI density \(n_{\rm CI}\) and inversely proportional to the
effective mass \(m^*\).

\section{Acoustic Phonon Scattering}

AP constitute an important intrinsic scattering mechanism
in BLG, particularly at finite lattice temperature.\cite{hwang2008acoustic,
arshia2021inelastic}
Here we consider longitudinal APs and evaluate the
energy-dependent carrier relaxation rate both with and without PB.
The analytical results are subsequently compared with numerical
calculations.

For longitudinal APs, a linear dispersion relation is assumed,
\(\omega_q=v_{lb}q,\)
where $v_{lb}$ is the longitudinal AP velocity and $q$ is
the phonon wave vector.\cite{hwang2008acoustic,arshia2021inelastic}

Using the acoustic deformation-potential coupling from
Eq.~(\ref{eq:coupling_functions}),
\(\frac{V_a^2\hbar q}{2A\rho_b v_{lb}}u_{kk'},
\)
and the BLG chiral overlap factor,\cite{mccann2013electronic,das2010theory}
the transport relaxation rate Eq.~(\ref{eq:rel_rate}), including PB (with \(f(x)=(1+exp((x-\mu)/K_B T))^{-1} \ ,\mu \ \rightarrow \) chemical potential),
is written as

\begin{multline}
\frac{1}{\tau_{\rm AP}^{\rm PB}}
=
\frac{V_a^2m^*}
{4\pi\hbar^2\rho_bv_{lb}^2}
\int_0^\infty\int_0^{2\pi}
\frac{1-f(E_{k'})}
{1-f(E_k)}
\\
\times
\omega_{ba}\left[
n_{\rm ph}
\delta(E_k+\hbar\omega_q-E_{k'})
+
(n_{\rm ph}+1)
\delta(E_k-\hbar\omega_q-E_{k'})
\right]
\\
\times
\frac{1+\cos(2\theta_{kk'})}{2}
(1-\cos\theta_{kk'})
\,d\theta_{kk'}\,dE_{k'} .
\label{eq:APSEWF}
\end{multline}

The resulting expression follows from the Fermi--Dirac occupation
factors and the AP transition probabilities within the
Boltzmann transport framework.\cite{das2010theory,arshia2021inelastic}
Evaluating the energy integral and carrying out the angular integration
under the approximations adopted here gives

\begin{equation}
    \frac{1}{\tau_{\rm AP}^{\rm PB}}
=
V_a^2 m^* \omega_{ba} \frac{}
{} \frac{ n_{ph}(1-f(E_{k}+\hbar\ \omega_{ba})) +(n_{ph}+1)(1-f(E_{k}-\hbar\ \omega_{ba})) \theta (E_k-\hbar\ \omega_{ba})}{4 \rho_b v_{lb}^2 \hbar ^2(1-f(E_k))}.
\label{eq:APSSWF}
\end{equation}

Here \(\omega_{ba}=\frac{4\sqrt{2m^*E_k}v_{lb}}{\pi \hbar}\), \(n_{ph}=\frac{1}{exp(\hbar\ \omega_{ba}/(K_b T))-1}\) and $\Theta(x)$ denotes the Heaviside step function. Its appearance in
the emission contribution imposes the kinematic threshold required for
phonon emission.

The absence of the PB factor considerably simplifies the analytical
evaluation. The corresponding relaxation rate without PB
is

\begin{equation}
\frac{1}{\tau_{\rm AP}}
=
V_a^2 m^* \omega_{ba} \frac{ n_{ph} +(n_{ph}+1) \theta (E_k-\hbar\ \omega_{ba})}{4 \rho_b v_{lb}^2 \hbar ^2}.
\label{eq:APSS}
\end{equation}

Equations~(\ref{eq:APSSWF}) and (\ref{eq:APSS}) provide the analytical
AP relaxation rates with and without PB,
respectively. Their comparison isolates the effect of the electronic
occupation factors on AP scattering. The temperature
dependence arises from both the phonon occupation factors and, when PB
is retained, the Fermi--Dirac occupation factors. Increasing the lattice
temperature generally enhances the AP population and
therefore increases the available scattering phase space.

\section{Optical Phonon Scattering}
OP scattering provides an inelastic relaxation channel whose importance depends strongly on the carrier-energy and temperature regime and on the particular BLG phonon branches included.\cite{borysenko2011electron,laitinen2015coupling}
In contrast to APs, OPs have a finite characteristic energy and arise
from out-of-phase lattice vibrations. Within the present model, the
OP branch is treated as approximately dispersionless over
the relevant range of wave vectors, so that
\(\omega_q\simeq\omega_o,\)
where $\omega_o$ is the OP frequency.

The OP occupation is described by the Bose--Einstein
distribution,
\(N_o=\frac{1}
{\exp(\beta\hbar\omega_o)-1}.\)
Using the optical deformation-potential matrix element specified in
Eq.~(\ref{eq:coupling_functions}),
\(|v_{kq}|^2=
\frac{V_o^2\hbar}
{A\rho_b\omega_o}\,u_{kk'},
\)
\cite{borysenko2011electron}
together with the BLG chiral overlap factor, the transport relaxation
rate in Eq.~(\ref{eq:rel_rate}) gives the PB-inclusive OP
scattering rate
\begin{multline}
\frac{1}{\tau_{\rm OP}^{\rm PB}}
=
\frac{V_o^{2}m^*}{2\pi\hbar^{2}\rho_b\omega_o}
\int_{0}^{\infty}\!\int_{0}^{2\pi}
\frac{1-f(E_{k'})}
     {1-f(E_k)}
\\
\times
\left[
N_o\,
\delta\!\left(E_k+\hbar\omega_o-E_{k'}\right)
+
(N_o+1)\,
\delta\!\left(E_k-\hbar\omega_o-E_{k'}\right)
\right]
\\
\times
\frac{1+\cos(2\theta_{kk'})}{2}
(1-\cos\theta_{kk'})
\,d\theta_{kk'}\,dE_{k'} .
\label{eq:OPSE_PB}
\end{multline}

Carrying out the angular integration using
\(
\int_0^{2\pi}
\frac{1+\cos(2\theta)}{2}
(1-\cos\theta)\,d\theta=\pi,
\) 
and using
\(k'dk'=\frac{m^*}{\hbar^2}dE_{k'},\)
the energy integration over the absorption and emission delta
functions yields

\begin{equation}
\frac{1}{\tau_{\rm OP}^{\rm PB}}
=
\frac{V_o^2m^*}
{2\rho_b\omega_o\hbar^2}
\left[
N_o
\frac{1-f(E_k+\hbar\omega_o)}
     {1-f(E_k)}
+
(N_o+1)
\Theta(E_k-\hbar\omega_o)
\frac{1-f(E_k-\hbar\omega_o)}
     {1-f(E_k)}
\right].
\label{eq:OPSS_PB}
\end{equation}
To isolate the effect of PB, the corresponding rate is also
evaluated by setting the occupation-factor ratio to unity. The corresponding analytical expression is

\begin{equation}
\frac{1}{\tau_{\rm OP}}
=
\frac{
V_o^2m^*
\left[
e^{\frac{\hbar\omega_o}{k_BT}}
\Theta(E_k-\hbar\omega_o)+1
\right]
}{
2\rho_b\omega_o\hbar^2
\left(
e^{\frac{\hbar\omega_o}{k_BT}}-1
\right)
}.
\label{eq:OPSS_noPB}
\end{equation}
Equation~(\ref{eq:OPSS_noPB}) shows that, in the absence of PB, the
OP scattering rate is determined by the phonon occupation
and the kinematic threshold for phonon emission.\cite{laitinen2015coupling}
The factor $\Theta(E_k-\hbar\omega_o)$ ensures that emission is allowed
only when $E_k\geq\hbar\omega_o$, whereas absorption remains possible
at all carrier energies provided that thermally populated OPs are available. Thus, within the present dispersionless
single-mode approximation, OP scattering possesses a
finite-energy emission threshold set by $\hbar\omega_o$.

\section{Surface Polar Phonon Scattering}

SPPs are optical vibrational modes localized
near the surface of polar dielectric substrates. Their relative
displacement of positively and negatively charged ions generates
macroscopic electric fields that couple to charge carriers through the
long-range Fr\"ohlich interaction. Because these evanescent electric
fields penetrate into the adjacent graphene layers, SPP scattering
provides an important channel for carrier momentum and energy relaxation,
particularly at elevated temperatures.\cite{fratini2008substrate,
konar2010effect,li2010surface}

Electron scattering by SPPs has been extensively
investigated in MLG on polar substrates
\cite{arshia2021inelastic}.
Although the underlying Boltzmann transport formalism is unchanged,
its application to BLG requires the appropriate parabolic dispersion,
DOS, chiral overlap factor, and dielectric environment
to be taken into account.\cite{das2010theory,li2011electron}

 Assuming that the carrier density is equally distributed between the
two graphene layers, the SPP transport relaxation rate is written as
\begin{multline}
\frac{1}{\tau_{SP}^{PB}}
=
\frac{m^*V_se_c^2}{2\pi\hbar^3}
\int_0^\infty\!\int_0^{2\pi}
\frac{1-f(E_{k'})}
     {1-f(E_k)}
\frac{
e^{-2qb}+e^{-2q(b+l)}
}{2q}
\\
\times
\left[
n_{ph}
\delta\!\left(E_k+\hbar\omega_s-E_{k'}\right)
+
(n_{ph}+1)
\delta\!\left(E_k-\hbar\omega_s-E_{k'}\right)
\right]
\\
\times
\frac{1+\cos(2\theta_{kk'})}{2}
(1-\cos\theta)
\,d\theta_{kk'}\,dE_{k'},
\label{eq:SPPSEWF}
\end{multline}
where
\(n_{ph}
=
\frac{1}
{\exp(\hbar\omega_s/k_BT)-1}
\)
is the Bose--Einstein occupation number of the SPP mode,
$\omega_s$ is the SPP frequency, and $b$ and $b+l$ characterize the
distances of the two graphene layers from the dielectric interface.

The first and second terms in Eq.~(\ref{eq:SPPSEWF}) describe phonon
absorption and emission, respectively. In particular, the emission
process is kinematically allowed only when
\(E_k\geq\hbar\omega_s.\)

Performing the energy and angular integrations analytically gives

\begin{equation}
\frac{1}{\tau_{SP}^{PB}}
=
\frac{m^{*}\,V_se_c^{\,2}}
{\sqrt{2}\,\pi\hbar^{2}}
\left[
n_{ph}T_{f+}
+
(n_{ph}+1)T_{f-}\Theta(E_k-\hbar\omega_s)
\right],
\label{eq:SPPSSWF}
\end{equation}

where the functions $T_{f+}$ and $T_{f-}$ contain the carrier-energy
dependence associated with phonon absorption and emission, respectively,
including the PB factor.

{\footnotesize
\begin{equation}
\begin{aligned}
T_{f\pm}
={}&
\frac{\mathscr{F_{\pm}}
}
{
30 \sqrt{2} \pi  \alpha_{\pm} ^3 \hbar ^2 \sqrt{ \left(2 \alpha _{\pm}+\beta_{\pm}\right)}
}
\\[1ex]
&\times
\Bigg[
\sqrt{\beta_{\pm}^2-4\alpha_{\pm}^2} \ \big(-17\alpha_{\pm}^2 -2\hbar^2 \omega_s^2 +5\alpha_{\pm} \beta_{\pm}\big)E[A_{\pm}]-\ \big(14E_k \alpha_{\pm}^2 + 11E_k \hbar^2 \omega_s^2 
\\[1ex]
&\qquad
{\pm}\hbar \omega_s (7E_k^2 + 2\hbar^2 \omega_s^2) +14\alpha_{\pm}^3 \ -\alpha_{\pm}\hbar^2\omega^2\big)E[B_{\pm}]
\\[1ex]
&\qquad
+ \sqrt{\beta_{\pm}^2-4\alpha_{\pm}^2} \ \big(13\alpha_{\pm}^2  +2\hbar^2 \omega_s^2 - 2E_k \alpha_{\pm}-\alpha_{\pm} \hbar \omega_s\big)K[A_{\pm}]
\\[1ex]
&\qquad
+
(30E_k^3 + 45E_k^2\hbar \omega_s + 19E_k \hbar^2 \omega_s^2 + 2\hbar^3 \omega_s^3 - 30\alpha_{\pm}^3 - 5\alpha \hbar^2 \omega_s^2)K[B_{\pm}]
\Bigg].
\end{aligned}
\label{eq:Tfpm}
\end{equation}
}

Here, $E(\cdot)$ and $K(\cdot)$ denote the complete elliptic integrals
of the second and first kinds, respectively, with arguments
\(
A_{\pm}
=
\frac{-4\sqrt{E_k(E_k\pm\hbar\omega_s)}}
{2E_k\pm\hbar\omega_s
-2\sqrt{E_k(E_k\pm\hbar\omega_s)}},\)
\(B_{\pm}
=
\frac{4\sqrt{E_k(E_k\pm\hbar\omega_s)}}
{2E_k\pm\hbar\omega_s
+2\sqrt{E_k(E_k\pm\hbar\omega_s)}},\)
\(\alpha_{\pm}
=\sqrt{E_k(E_k\pm\hbar\omega_s}\),
 \(\beta_{\pm}
=
2E_k\pm\hbar\omega_s.\) and \(\mathscr{F_{\pm}}=\frac{
1-F_{db}(E_k\pm\hbar\omega_s)
}
{1-F_{db}(E_k)}\)

For the emission contribution, the expressions containing the minus
sign are understood only for
$E_k\geq\hbar\omega_s$, or equivalently are multiplied by
$\Theta(E_k-\hbar\omega_s)$.

To isolate the effect of PB, we also consider the
approximation 
\(\frac{1-f(E_{k'})}{1-f(E_k)}\simeq1.\)
Under this approximation, Eq.~(\ref{eq:SPPSEWF}) becomes

\begin{equation}
\frac{1}{\tau_{SP}}
=
\frac{V_se_c^2m^*}
{\sqrt{2}\pi\hbar^2}
\left[
n_{ph}T_+
+
(n_{ph}+1)T_-\Theta(E_k-\hbar\omega_s)
\right],
\label{eq:SPPSS}
\end{equation}

where $T_{\pm}$ are obtained from $T_{f\pm}$ by removing the
PB ratio (\(\mathscr{F_{\pm}}=1\)). The emission term remains subject to the
kinematic condition
$E_k\geq\hbar\omega_s$.

\section{Results and Discussion}

The transport relaxation rate is one of the key quantities governing
electronic transport in BLG. It is strongly influenced by various
scattering mechanisms, including NI, SR, CI, AP, OP,
and SPP scatterings.
Understanding the relative contributions of these mechanisms is essential
for assessing and optimizing the transport performance of BLG-based
electronic devices.\cite{das2010theory,li2011electron,xiao2010charged,
hwang2008acoustic,fratini2008substrate,li2010surface}

In this section, the energy-dependent transport relaxation rates
corresponding to the individual scattering mechanisms are presented and
discussed. Unless stated otherwise, the carrier energy is varied over
the range
\(0\leq E_k\leq0.35~\mathrm{eV}.\) The analytical expressions derived in the preceding sections are
evaluated numerically, and the resulting rates are verified by direct
numerical evaluation of the corresponding transport integrals.

The principal parameters used in the numerical calculations are
summarized in Table~\ref{tab:parameters}. The effective mass follows the standard low-energy
effective-mass description of Bernal-stacked BLG.\cite{mccann2013electronic}

\begin{table*}[h!]
\centering
\renewcommand{\arraystretch}{1.2}
\small

\begin{tabular}{|p{1.2cm}|p{6cm}|p{4.5cm}|}
\hline
\textbf{Symbol} & \textbf{Physical meaning} & \textbf{Value} \\
\hline

$\hbar$ & Reduced Planck constant &
$1.05457\times10^{-27}\,\mathrm{erg\,s}$ \\
\hline

$A$ & Area of the BLG sheet &
Defined by sample geometry \\
\hline

$m^{*}$ & Effective mass of electrons in BLG &
$0.033\,m_e$ \\
\hline

$k_F$ & Fermi wave vector &
Calculated from $n$ \\
\hline

$n$ & Carrier concentration &
$1\times10^{12}\,\mathrm{cm^{-2}}$ \\
\hline

$e_c$ & Elementary charge in Gaussian units &
$4.8032\times10^{-10}\,\mathrm{esu}$ \\
\hline

$k_B$ & Boltzmann constant &
$1.38065\times10^{-16}\,\mathrm{erg\,K^{-1}}$ \\
\hline

$\rho_b$ & Areal mass density of BLG &
$15.2\times10^{-8}\,\mathrm{g\,cm^{-2}}$ \\
\hline

$v_{lb}$ & Longitudinal acoustic phonon velocity &
$2.12\times10^{6}\,\mathrm{cm\,s^{-1}}$
\cite{hwang2008acoustic} \\
\hline

$V_0$ & Neutral impurity scattering potential &
$1.0\,\mathrm{eV}$ \\
\hline

$a$ & Neutral impurity correlation length &
$0.5\,\mathrm{nm}$ \\
\hline

$n_{\rm CI}$ & Charged impurity concentration &
$1\times10^{10}\,\mathrm{cm^{-2}}$ \\
\hline

$n_{\rm int}$ & Interface charge density &
$1\times10^{10}\,\mathrm{cm^{-2}}$ \\
\hline

$V_a$ & Acoustic deformation potential &
$19\,\mathrm{eV}$ \\
\hline

$V_o$ & Optical deformation potential &
$2.5\,\mathrm{eV\,\AA^{-1}}$ \\
\hline

$V_s$ & Surface polar phonon coupling parameter &
Given by Eq.~(\ref{eq:vs}) \\
\hline

$\omega_q$ & Acoustic phonon dispersion &
$\omega_q=v_{lb}q$ \\
\hline

$\hbar\omega_o$ & Optical phonon energy &
$200\,\mathrm{meV}$ \\
\hline

$\hbar\omega_{sp}$ & Surface polar phonon energy &
$59.8\,\mathrm{meV}$ (SiO$_2$ substrate) \\
\hline

$\epsilon_s$ & Effective background dielectric constant &
$2.45$ \\
\hline

$\epsilon_l$ & Low-frequency relative dielectric constant &
$3.9$ \\
\hline

$\epsilon_h$ & High-frequency relative dielectric constant &
$2.5$ \\
\hline

$\epsilon_q$ & Static dielectric screening function &
Given by Eq.~(\ref{eq:dielectric}) \\
\hline

$b$ & Distance between the substrate and BLG &
$50\,\mathrm{\AA}$ \\
\hline

$l$ & Interlayer spacing of BLG &
$3.37\,\mathrm{\AA}$ \\
\hline

$d$ & Surface-roughness correlation length &
$100\,\mathrm{\AA}$ \\
\hline

$h$ & Surface-roughness amplitude &
$5\,\mathrm{\AA}$ \\
\hline

$T$ & Temperature used in the calculations &
$150\,\mathrm{K}$(AP);
$200\,\mathrm{K}$(OP and SPP) \\
\hline

\end{tabular}

\caption{Material parameters and symbols used in the scattering
formalism. Numerical parameters correspond to the values adopted in
the present calculations. Parameters without numerical values are
defined symbolically and evaluated through the corresponding
equations.}
\label{tab:parameters}
\end{table*}

\subsection{Neutral Impurity Scattering}

Unless otherwise stated, the numerical calculations for NI
scattering use a NI density
\(
n_{\rm NI}=10^{11}\,\mathrm{cm^{-2}},
\)
a Gaussian correlation length
\(
a=0.5\,\mathrm{nm},
\)
and an impurity potential amplitude
\(
V_0=1.0\,\mathrm{eV}.
\)
The parameters \(a\) and \(V_0\) are phenomenological quantities used to
characterize the spatial extent and strength of finite-range neutral
disorder, respectively. Finite-range disorder models are commonly used
in theoretical treatments of graphene transport.\cite{lewenkopf2013computational,ortmann2015graphene}
The values adopted here are representative simulation parameters rather
than unique material constants.

Using Eq.~(\ref{eq:tau_final}), the energy dependence of the NI transport relaxation rate is evaluated numerically. The
result is compared with a direct numerical evaluation of the angular
integral in Eq.~(\ref{eq:tau_angle}) to verify the analytical
Bessel-function expression.

To place the analytical BLG result in a broader context, we compare
finite-range NI scattering in BLG with the corresponding
results for MLG and a conventional 2DEG. Within the same first Born approximation and
Gaussian impurity model, the differences among these systems arise from
their electronic band structures, DOS, and angular
overlap factors.\cite{ando1982electronic,adam2011graphene,mccann2013electronic}

For all three systems, the transport relaxation rate can be written in
the unified form
\begin{equation}
\frac{1}{\tau_{NI}}
=
\frac{2\pi n_{\rm NI}}{\hbar}
(2\pi a^2V_0)^2
D(E_F)
R(x),
\qquad
x=2k_F^2a^2,
\label{eq:unified_NI}
\end{equation}
where \(D(E_F)\) is the DOS appropriate to the system
under consideration and \(R(x)\) is a dimensionless transport kernel
containing the angular dependence of the scattering process.

The three systems differ in their angular overlap factors. For a
conventional 2DEG,
\(
u_{\rm 2DEG}=1,
\)
whereas for MLG,
\(
u_{\rm MLG}
=
\frac{1+\cos\theta}{2}.
\)
These different overlap factors lead to distinct angular weightings of
the transport scattering probability.\cite{castro2009electronic,adam2011graphene,mccann2013electronic}

\begin{equation}
\begin{aligned}
R_{\rm 2DEG}(x)
&=
e^{-x}
\left[
I_0(x)-I_1(x)
\right],
\\[2mm]
R_{\rm MLG}(x)
&=
\frac{1}{4}e^{-x}
\left[
I_0(x)-I_2(x)
\right],
\\[2mm]
R_{\rm BLG}(x)
&=
e^{-x}
\left[
\frac{1}{2}I_0(x)
-\frac{3}{4}I_1(x)
+\frac{1}{2}I_2(x)
-\frac{1}{4}I_3(x)
\right].
\end{aligned}
\label{eq:comparison}
\end{equation}

For MLG, Eq.~(\ref{eq:unified_NI}) using (\ref{eq:comparison}) gives
\begin{equation}
\frac{1}{\tau_{NI}^{\rm MLG}}
=
\frac{2\pi n_{\rm NI}}{\hbar}
(2\pi a^2V_0)^2
D_{\rm MLG}(E_F)
e^{-x}
\frac{1}{4}
\left[
I_0(x)-I_2(x)
\right].
\label{eq:tau_MLG}
\end{equation}

Using the modified-Bessel-function identity
\(I_0(x)-I_2(x)
=
\frac{2I_1(x)}{x},
\) this may equivalently be written as
\begin{equation}
\frac{1}{\tau_{NI}^{\rm MLG}}
=
\frac{2\pi n_{\rm NI}}{\hbar}
(2\pi a^2V_0)^2
D_{\rm MLG}(E_F)
e^{-x}
\frac{I_1(x)}{2x}.
\label{eq:tau_MLG_alternative}
\end{equation}
This form is consistent with the general Boltzmann transport treatment
of disorder in graphene.\cite{adam2011graphene}
An additional distinction among the three systems is the energy
dependence of the DOS. For a conventional parabolic 2DEG,
\(
D_{\rm 2DEG}(E)
=
\frac{g_s m_{\rm 2DEG}^*}{2\pi\hbar^2},
\)
whereas for low-energy BLG,
\(
D_{\rm BLG}(E)
=
\frac{g_sg_vm_{\rm BLG}^*}{2\pi\hbar^2}.
\)
Thus, both the conventional 2DEG and low-energy BLG possess
energy-independent DOS, although their numerical values
depend on their respective effective masses and degeneracy factors. In
contrast, the MLG DOS varies linearly with energy.\cite{ando1982electronic,castro2009electronic,mccann2013electronic}

In the short-range limit, \(x\rightarrow0\), the transport kernels
reduce to
\(
R_{\rm 2DEG}(0)=1,
\quad
R_{\rm BLG}(0)=\frac{1}{2},
\quad
R_{\rm MLG}(0)=\frac{1}{4}.
\)
These differences reflect the distinct angular weighting introduced by
the overlap factors of the three systems. The comparison between the
2DEG and BLG is particularly useful because both possess parabolic
low-energy dispersions and energy-independent DOS,
whereas their angular scattering structures differ because BLG carries
a nontrivial pseudospin chirality. MLG introduces the additional
distinction of a linear Dirac dispersion and an energy-dependent DOS.

Thus, Eq.~(\ref{eq:comparison}) provides a compact unified description
of finite-range NI scattering in 2DEG, MLG, and BLG.
The 2DEG and MLG transport descriptions are consistent with established
Boltzmann treatments,\cite{ando1982electronic,adam2011graphene}
while the BLG kernel follows directly from the low-energy two-band
chiral overlap factor and the analytical Gaussian-impurity calculation
developed in the present work.

The energy dependence of the calculated scattering rates is shown in
Fig.~\ref{fig:DMB}. Panel~(a) compares the SR, NI, and CI scattering
rates in MLG and BLG, whereas panel~(b) examines the effect of dielectric
screening on CI and SR scattering in BLG. The subscripts \(m\) and \(b\)
denote monolayer and bilayer graphene, respectively.
Figure~\ref{fig:DMB}(a) shows a pronounced difference between the
NI scattering rates of BLG and MLG. For BLG, the
NI rate, $NI_b$, is the dominant scattering contribution
over most of the investigated energy range, although the
SR contribution is larger in the very-low-energy
regime. The rate $NI_b$ is of the order of
$10^{12}$--$10^{13}~\mathrm{s^{-1}}$ and exhibits only a weak
dependence on carrier energy. This behavior is consistent with the
approximately energy-independent DOS of low-energy BLG.
Thus, within the present low-energy approximation, the energy
dependence of the BLG NI scattering rate arises
primarily through the finite-range transport kernel, since the BLG
DOS is approximately energy independent.

In contrast, the MLG NI rate, $NI_m$, is substantially
smaller at low carrier energies and increases gradually with
increasing $E_k$, reaching approximately
$10^{10}$--$10^{11}~\mathrm{s^{-1}}$ toward the upper part of the
investigated energy range. This behavior reflects the combined
influence of the linearly increasing MLG DOS and the
energy dependence of the finite-range transport kernel. Consequently, the calculated NI rates
exhibit markedly different energy dependencies in MLG and BLG.

The results shown in Fig.~\ref{fig:DMB}(a) therefore provide a clear
numerical manifestation of the different electronic structures of
MLG and BLG and their influence on NI-limited carrier
transport.

It should be noted that a separately screened NI scattering case has not
been considered. Unlike CI, NI do not
generate a long-range Coulomb potential, and their scattering potential
is predominantly short-ranged. Hence, the long-range dielectric
screening treatment used for CI scattering does not lead to an analogous
screening correction for the NI potential adopted here. 
\subsection{Surface Roughness Scattering}

The SR model adopted here is based on the
interface-roughness framework discussed in
Refs.~\cite{shishir2009room,shah2024role}, with the corresponding
formulation generalized in the present work to the low-energy BLG
electronic structure. The complete screened SR scattering rate is given
by Eq.~(\ref{eq:SRSR}).

To verify the analytical result, the corresponding unscreened transport
integral was also evaluated numerically using
Eq.~(\ref{eq:SRS_angular}). The analytical and direct numerical results
show close agreement over the investigated energy range, confirming that
the closed-form expression in Eq.~(\ref{eq:SRS_analytical}) correctly
reproduces the corresponding transport integral.
The analytical unscreened SR scattering rate obtained from
Eq.~(\ref{eq:SRS_analytical}) is shown in Fig.~\ref{fig:DMB}(a) as a
function of carrier energy. For BLG, the SR scattering rate decreases
strongly with increasing carrier energy, with the most pronounced
variation occurring in the low-energy region below approximately
\(0.1~\mathrm{eV}\). At higher carrier energies, the rate decreases more
gradually, where the impurity
scattering contributions, NI and CI become comparatively more important. This behavior results from the combined energy dependence of
the Fermi wave vector, the Gaussian roughness factor, and the transport
angular kernel. In particular, increasing carrier energy increases
\(k_F\), thereby modifying the dimensionless parameter
\(
x_{\rm SR}=\frac{k_F^2d^2}{2},
\)
and consequently the Bessel-function combination appearing in
Eq.~(\ref{eq:SRS_analytical}). Thus, the energy dependence arises from
the complete analytical expression rather than from the Gaussian
roughness factor alone.

The effect of dielectric screening on SR scattering is examined in
Fig.~\ref{fig:DMB}(b), which compares the SR rates obtained without
screening, with the full momentum-dependent screening, and within the
asymptotic strong-screening approximation. In the present model,
screening enters the SR scattering potential through the electrostatic
response of the graphene--dielectric interface. The fully screened
result is obtained from Eq.~(\ref{eq:SRSR}), whereas the unscreened
result corresponds to Eq.~(\ref{eq:SRS_analytical}). The screening
factor, appearing through \(\epsilon_q^{-2}\), modifies the effective
scattering potential and therefore the magnitude and energy dependence
of the SR relaxation rate.

As shown in Fig.~\ref{fig:DMB}(b), the unscreened SR rate,
\(\mathrm{SR}_{w}\), is substantially larger than the screened SR rates
over most of the investigated energy range. The fully screened rate,
\(\mathrm{SR}_{f}\), exhibits a pronounced low-energy maximum before
decreasing with increasing carrier energy. In contrast, the
strong-screening approximation, \(\mathrm{SR}_{s}\), substantially
suppresses the SR rate and produces a much smaller rate over the
investigated range. Thus, the treatment of dielectric screening has a
significant influence not only on the magnitude of SR scattering but
also on its energy dependence.

A direct comparison of SR and CI scattering in Fig.~\ref{fig:DMB}(b)
further illustrates the role of screening in BLG. In the absence of
screening, both mechanisms produce relatively large relaxation rates,
with the unscreened SR rate becoming particularly prominent toward the
higher-energy region. When full screening is included, the SR rate
exhibits a low-energy maximum and subsequently decreases, while the
screened CI rate remains comparatively large over a broader energy
range. Under the strong-screening approximation, the SR rate is strongly
suppressed and remains below the corresponding screened CI rate over
most of the investigated energy range.

These results demonstrate that the relative importance of SR and CI
scattering in BLG depends sensitively on both carrier energy and the
treatment of dielectric screening. In particular, screening changes the
balance between the two mechanisms rather than producing a uniform
rescaling of their scattering rates. The calculated results therefore
indicate that dielectric screening should be included when assessing the
relative contributions of SR and CI scattering in BLG.

\subsection{Charged Impurity Scattering}

CI scattering is an important mechanism that
contributes to the degradation of carrier mobility and conductivity in
BLG. CIs associated with substrate defects, adsorbed
ions, and fabrication-induced contaminants generate long-range Coulomb
potentials and can therefore provide a significant source of momentum
relaxation.\cite{das2010theory,xiao2010charged,adam2011graphene}

Under the approximation
\(e^{-2qb}\simeq e^{-2q(b+l)}\simeq 1,\)
corresponding to impurity distances satisfying
\(qb\ll1\) and \(q(b+l)\ll1\), the analytical CI relaxation rate is
obtained as given in Eq.~(\ref{eq:CISE}).

The complete expression in Eq.~(\ref{eq:CI_integral}), without the
zero-separation approximation, is evaluated numerically. This provides
a direct assessment of the validity and range of applicability of the
analytical approximation. The analytical and numerical results show
similar energy dependence and are found to be in close agreement. The
remaining differences in magnitude arise primarily from the treatment
of the finite-distance impurity factors.

The calculated CI relaxation rate exhibits a stronger energy dependence
in the low-energy region. This behavior is associated with the energy
dependence of the carrier wave vector and the Thomas--Fermi screening
parameter. In low-energy BLG, the parabolic dispersion
\(E_k=\frac{\hbar^2k^2}{2m^*}\)
gives \(k\propto E_k^{1/2}\), whereas the Thomas--Fermi screening wave
vector \(q_{\rm TF}\) is approximately energy independent. Consequently,
\(\alpha=\frac{q_{\rm TF}}{2k}\propto E_k^{-1/2}.\)
Thus, screening becomes increasingly important at low carrier energies. When the numerical calculation
is performed using discrete energy points, the apparent discreteness of
the plotted rate reflects the chosen energy sampling and should not be
interpreted as evidence that only a limited number of physical
electronic states satisfy energy conservation.

The dependence of the CI relaxation rate on impurity density is examined
numerically. Within the independent-scatterer approximation, the
transport rate is proportional to the CI density,
\(\frac{1}{\tau_{\rm CI}}\propto n_{\rm CI}.\)
This proportionality is the standard result for independent charged
scatterers in transport theory.\cite{dassarma2011electronic,xiao2010charged}

The CI scattering characteristics of BLG can be placed in a broader
context by comparing them with those of MLG and a conventional 2DEG.
The three systems differ in their low-energy dispersions, DOS, screening properties, and wave-function overlap
factors.\cite{ando1982electronic,castro2009electronic,
dassarma2011electronic,mccann2013electronic}

Within the Thomas--Fermi approximation, the effective coupling function
for screened Coulomb scattering in MLG is
\(
v^{\rm MLG}_{kk'}
=
\frac{2\pi e_c^2}
{\epsilon_s(q+q_{\rm TF}^{\rm MLG})}
e^{-qb}u_{kk'}^{\rm MLG},
\)
where
\(
q_{\rm TF}^{\rm MLG}=4r_sk_F,
\qquad
r_s=\frac{e_c^2}{\epsilon_s\hbar v_F}.
\)
Thus,
\(
q_{\rm TF}^{\rm MLG}\propto k_F\propto\sqrt{n},
\)
while
\(
\frac{q_{\rm TF}^{\rm MLG}}{k_F}=4r_s
\)
is independent of carrier density for a fixed dielectric environment.
The result \(q_{\rm TF}^{\rm MLG}=4r_sk_F\) is standard for doped
MLG.\cite{hwang2009screening,dassarma2011electronic}

For low-energy BLG, the Thomas--Fermi screening wave vector is
\(
q_{\rm TF}^{\rm BLG}
=
\frac{4m^*e_c^2}{\epsilon_s\hbar^2},
\)
for spin and valley degeneracy \(g_sg_v=4\).
\cite{das2010theory,mccann2013electronic}
Thus, within the parabolic low-energy approximation,
\(q_{\rm TF}^{\rm BLG}\) is independent of carrier density.

For a conventional parabolic 2DEG with total degeneracy \(g\),
\(
q_{\rm TF}^{\rm 2DEG}
=
\frac{g m_{\rm 2DEG}^*e_c^2}
{\epsilon_s\hbar^2},
\)
which is likewise independent of carrier density within the ideal
parabolic-band approximation. For a spin-degenerate 2DEG, \(g=2\).
\cite{ando1982electronic,dassarma1985relaxation}

The principal differences among the three systems are summarized in
Table~\ref{tab:MLG_BLG_2DEG_comparison}. The electronic structures and
chiral overlap factors of MLG and BLG are well established,
\cite{castro2009electronic,mccann2013electronic}
whereas the corresponding parabolic 2DEG results follow the standard
two-dimensional electron-gas treatment.\cite{ando1982electronic}

The comparison shows that low-energy BLG shares the parabolic
dispersion, approximately constant DOS, and density-independent
Thomas--Fermi screening of a conventional 2DEG, while retaining the
chiral pseudospin structure of graphene. In particular,
\(
u_{\rm MLG}(\pi)=0,
\qquad
u_{\rm BLG}(\pi)=1,
\qquad
u_{\rm 2DEG}(\pi)=1.
\)
Thus, MLG suppresses exact backscattering, whereas low-energy BLG does
not. Nevertheless, the angular dependence of BLG remains distinct from
that of a conventional 2DEG because of its chiral overlap factor.
\cite{castro2009electronic,mccann2013electronic}

For elastic scattering at the Fermi surface, the CI relaxation rate in
MLG is
\begin{equation}
\frac{1}{\tau_{\rm CI}^{\rm MLG}}
=
\frac{4\pi n_{\rm CI}r_s^2v_F}{k_F}
\int_0^1
\frac{x^2\sqrt{1-x^2}}
{(x+2r_s)^2}
e^{-4k_Fbx}\,dx,
\label{eq:MLG_CI_dimensionless}
\end{equation}
where
\(x=\sin\frac{\theta}{2}\)
and \(q=2k_Fx.
\)

For finite impurity distance \(b\), the factor \(e^{-4k_Fbx}\)
prevents the integral from reducing to the simple closed form obtained
in the \(b=0\) limit. Although formal representations in terms of
special functions or convergent series can be constructed, they are
less convenient for numerical evaluation. Therefore, the exact
finite-\(b\) integral in Eq.~(\ref{eq:MLG_CI_dimensionless}) is evaluated
numerically, while the \(b=0\) limit is treated analytically.

For impurities located in the graphene plane (\(b=0\)), the remaining
integral can be evaluated analytically. For \(r_s>1/2\), one obtains
\begin{equation}
\begin{aligned}
\frac{1}{\tau_{\rm CI}^{\rm MLG}}
=
\frac{4\pi n_{\rm CI}r_s^2v_F}{k_F}
\Bigg[
&\frac{\pi}{4}
+6r_s
-6\pi r_s^2
\\
&+
\frac{8r_s(6r_s^2-1)}
{\sqrt{4r_s^2-1}}
\left\{
\arctan\left(
\frac{2r_s+1}{\sqrt{4r_s^2-1}}
\right)
-
\arctan\left(
\frac{1}{\sqrt{4r_s^2-1}}
\right)
\right\}
\Bigg].
\label{eq:MLG_CI_final_bzero}
\end{aligned}
\end{equation}

For impurities located in the 2DEG plane (\(b=0\)), the corresponding
integral can also be evaluated analytically. For
\(r_s^{\rm 2DEG}>1\),
\begin{equation}
\frac{1}{\tau_{\rm CI}^{\rm 2DEG}}
=
\frac{4\pi m_{\rm 2DEG}^* n_{\rm CI}e_c^4}
{\hbar^3\epsilon_s^2 k_F^2}
\left[
\frac{\pi}{2}
-\frac{r_s^{\rm 2DEG}}
{(r_s^{\rm 2DEG})^2-1}
+
\frac{r_s^{\rm 2DEG}
\left[2-(r_s^{\rm 2DEG})^2\right]}
{\left[(r_s^{\rm 2DEG})^2-1\right]^{3/2}}
\arccos\left(
\frac{1}{r_s^{\rm 2DEG}}
\right)
\right].
\label{eq:2DEG_CI_final_bzero}
\end{equation}

For finite impurity distance \(b\), the corresponding integral is
evaluated numerically using the 2DEG DOS. Unlike MLG and BLG, no chiral
wave-function overlap factor appears in the conventional 2DEG transport
integral.

Overall, CI scattering in BLG combines the approximately
density-independent screening characteristic of a parabolic
two-dimensional system with a nontrivial graphene pseudospin overlap,
resulting in transport characteristics distinct from those of both MLG
and a conventional 2DEG.
\cite{das2010theory,mccann2013electronic,ando1982electronic}

\begin{figure}[ht!]
\centering
\begin{subfigure}[b]{0.45\textwidth}
\centering
\includegraphics[width=\textwidth]{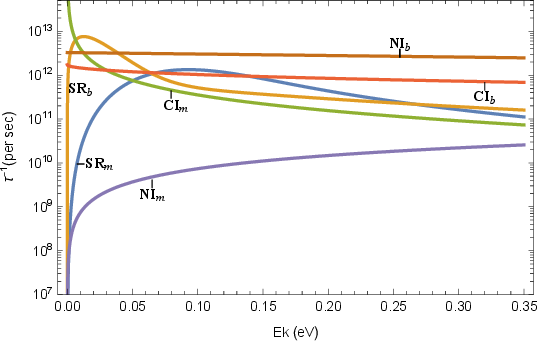}
\caption{}
\end{subfigure}
\hfill
\begin{subfigure}[b]{0.45\textwidth}
\centering
\includegraphics[width=\textwidth]{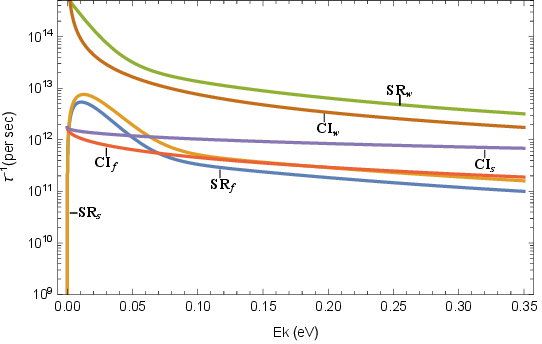}
\caption{}
\end{subfigure}
\caption{(a) Energy dependence of the carrier scattering rates,
$\tau^{-1}$, in monolayer (subscript $m$) and bilayer (subscript $b$)
graphene due to surface-roughness (SR), neutral-impurity (NI), and
charged-impurity (CI) scattering. (b) Energy dependence of the CI and
SR scattering rates in bilayer graphene for three screening regimes:
without screening ($w$), full screening ($f$), and the asymptotic
strong-screening approximation ($s$). The corresponding rates are
denoted by $\mathrm{CI}_{w}$, $\mathrm{CI}_{f}$, $\mathrm{CI}_{s}$ and
$\mathrm{SR}_{w}$, $\mathrm{SR}_{f}$, $\mathrm{SR}_{s}$.
}
\label{fig:DMB}
\end{figure}

In Fig.~\ref{fig:DMB}(a), the CI scattering rates exhibit markedly
different energy dependences in MLG and BLG. For BLG, the CI rate,
$\mathrm{CI}_b$, is relatively large at low carrier energy and
decreases moderately with increasing energy. It is of the order of
$10^{12}~\mathrm{s^{-1}}$ at low carrier energy and decreases to several
$10^{11}~\mathrm{s^{-1}}$ at
\(E_k\simeq0.35~\mathrm{eV}\). In contrast, the MLG CI rate,
$\mathrm{CI}_m$, exhibits a pronounced non-monotonic dependence. It
increases rapidly at very low energy, reaches a maximum of approximately
$10^{12}~\mathrm{s^{-1}}$ around
\(E_k\simeq0.08\)--\(0.10~\mathrm{eV}\), and subsequently decreases with
increasing carrier energy.

\begin{table}[htbp]
\centering
\renewcommand{\arraystretch}{1.2}
\begin{tabular}{c|c|c|c|c|c}
\hline
\textbf{System}
&
\textbf{Dispersion}
&
\textbf{$D(E_F)$}
&
\textbf{$q_{\rm TF}$}
&
\textbf{$u(\theta)$}
&
\textbf{$u(\pi)$}
\\
\hline

MLG
&
$E_k=\hbar v_F k$
&
$\dfrac{g_sg_vk_F}{2\pi\hbar v_F}$
&
$\propto k_F$
&
$\dfrac{1+\cos\theta}{2}$
&
$0$
\\[8pt]

BLG
&
$E_k=\dfrac{\hbar^2k^2}{2m^*}$
&
$\dfrac{g_sg_vm^*}{2\pi\hbar^2}$
&
constant
&
$\dfrac{1+\cos2\theta}{2}$
&
$1$
\\[8pt]

2DEG
&
$E_k=\dfrac{\hbar^2k^2}{2m_{\rm 2DEG}^*}$
&
$\dfrac{g_sm_{\rm 2DEG}^*}{2\pi\hbar^2}$
&
constant
&
$1$
&
$1$
\\
\hline
\end{tabular}
\caption{Comparison of the low-energy electronic properties relevant to
charged-impurity scattering in monolayer graphene (MLG), bilayer graphene
(BLG), and a conventional two-dimensional electron gas (2DEG).}
\label{tab:MLG_BLG_2DEG_comparison}
\end{table}

The different energy dependences of $\mathrm{CI}_m$ and
$\mathrm{CI}_b$ arise from the distinct electronic structures of MLG
and BLG. In MLG, the carrier wave vector, density of states, screening
wave vector, and chiral overlap factor all contribute to the energy
dependence of the scattering rate. In particular,
\(q_{\rm TF}^{\rm MLG}=4r_sk_F,\)
so that the screening wave vector varies with \(k_F\) and hence with
carrier energy. These combined dependencies produce the non-monotonic
behavior of $\mathrm{CI}_m$. In low-energy BLG, by contrast, the
approximately parabolic dispersion gives an approximately constant DOS
and a density-independent Thomas--Fermi screening wave vector. The
resulting CI rate therefore exhibits a smoother energy dependence.

Around \(E_k\simeq0.1~\mathrm{eV}\), the MLG and BLG CI rates become
comparable in magnitude. At higher carrier energies,
$\mathrm{CI}_b$ remains larger than $\mathrm{CI}_m$. Thus, the relative
importance of CI scattering in MLG and BLG is strongly energy dependent
and reflects the combined effects of their different dispersions, DOS,
screening properties, and chiral overlap factors.

The influence of dielectric screening on CI scattering is considered
by comparing screened and unscreened relaxation rates. Screening reduces
the strength of the long-range Coulomb interaction and consequently
modifies the magnitude of the CI relaxation rate, particularly at small
momentum transfer.\cite{das2010theory,adam2007self,xiao2010charged}
The comparison demonstrates that dielectric screening has a significant
effect on the calculated CI rate and should therefore be retained when
estimating realistic CI scattering in BLG rather than treating the
Coulomb interaction as unscreened.
The effect of dielectric screening on CI and SR scattering in BLG is
shown in Fig.~\ref{fig:DMB}(b). The unscreened CI rate,
$\mathrm{CI}_{w}$, is substantially larger than the corresponding
screened rates over the energy range considered. Inclusion of screening
therefore significantly suppresses the strength of the long-range
Coulomb interaction and also modifies the energy dependence of the CI
relaxation rate.\cite{HwangDasSarma2008}

\subsection{Acoustic Phonon Scattering}

\begin{figure}[ht]
\centering
\begin{subfigure}[b]{0.45\textwidth}
\centering
\includegraphics[width=\textwidth]{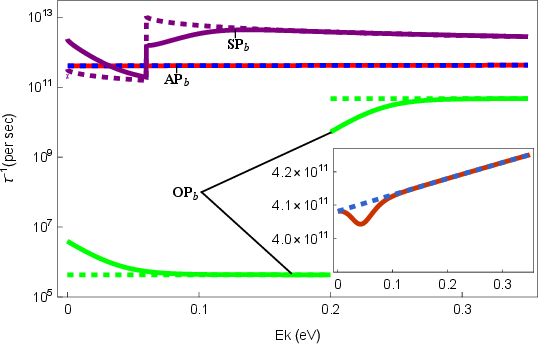}
\end{subfigure}
\hfill
\begin{subfigure}[b]{0.45\textwidth}
\centering
\includegraphics[width=\textwidth]{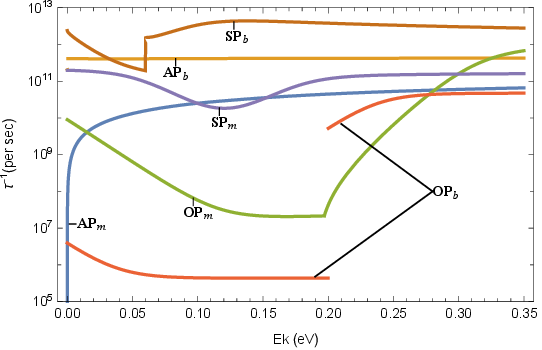}
\end{subfigure}
\caption{(a) Calculated carrier transport scattering rates,
$\tau^{-1}$, as a function of carrier energy $E_k$ in bilayer graphene
for AP (AP$_b$), optical-phonon (OP$_b$), and
surface-polar-phonon (SPP$_b$) scattering mechanisms. Solid curves
represent the results without Pauli-blocking (PB), whereas dashed curves
include the PB factor. The inset shows an enlarged
view of the acoustic-phonon contribution, highlighting its energy
dependence in the range of approximately $0$--$0.35~\mathrm{eV}$.
(b) Calculated carrier transport scattering rates,
$\tau^{-1}$, as a function of carrier energy $E_k$ for monolayer
and bilayer graphene due to acoustic-phonon (AP), optical-phonon (OP),
and surface-polar-phonon (SPP) scattering mechanisms. The subscripts
$m$ and $b$ denote monolayer and bilayer graphene, respectively. The
figure compares the energy dependence and relative magnitudes of the
phonon-limited scattering rates in the two graphene systems.}
\label{fig:PMBLG}
\end{figure}
The AP transport relaxation rate in BLG was evaluated both with and
without PB, using the analytical expressions derived in
Eqs.~(\ref{eq:APSSWF}) and (\ref{eq:APSS}), respectively. The
corresponding numerical transport integrals were evaluated
independently to assess the validity of the analytical results. The two
approaches coincide within numerical accuracy over the investigated
energy range. The calculated phonon-limited transport relaxation rates
in BLG are summarized in Fig.~\ref{fig:PMBLG}(a), which shows the
contributions from AP$_b$, OP$_b$, and SPP$_b$ scattering as functions
of carrier energy. The solid curves represent calculations without the
PB factor, whereas the dashed curves include PB. The separation
between the two curves therefore indicates the modification of the
transport relaxation rate due to final-state occupation. The
corresponding comparison between MLG and BLG is shown in
Fig.~\ref{fig:PMBLG}(b), where the subscripts $m$ and $b$ denote MLG
and BLG, respectively.

The calculated AP transport relaxation rate in BLG exhibits distinct
absorption and emission contributions. The absorption contribution
extends toward low carrier energies, whereas the finite-phonon-energy
treatment imposes a kinematic threshold on phonon emission, since the
carrier must possess sufficient energy to satisfy the emission
condition. Consequently, the available phase space for emission is
restricted near the threshold, giving rise to the pronounced
low-energy structure observed in the AP contribution in
Fig.~\ref{fig:PMBLG}(a). This behavior reflects the role of the
phonon phase space and finite phonon energy in determining the
energy dependence of AP transport relaxation.

The effect of PB is most pronounced in the low-energy region, where
the solid and dashed AP curves in Fig.~\ref{fig:PMBLG}(a) show a
noticeable separation. At higher carrier energies, the two curves
remain relatively close, indicating that PB has only a modest effect
on the AP transport relaxation rate over most of the investigated
range. The inset of Fig.~\ref{fig:PMBLG}(a) provides a direct
comparison of the AP transport relaxation rates calculated with and
without PB and confirms this trend. The enhanced low-energy
difference reflects the stronger restriction of available final
electronic states by Pauli exclusion. Thus, while PB does not
substantially alter the AP transport relaxation rate over the full
energy range, its inclusion is important for an accurate description
of the low-energy scattering regime.

For comparison, the AP relaxation rate in BLG is
considered alongside the corresponding behavior in MLG and a
conventional 2DEG. The calculated MLG and BLG AP rates
are shown by the $AP_m$ and $AP_b$ curves, respectively, in
Fig.~\ref{fig:PMBLG}(b). In the high-temperature equipartition regime,
where
\(\hbar\omega_{\rm AP}\ll k_BT,\)
AP scattering may be treated in the quasi-elastic
approximation.\cite{hwang2008acoustic}

For MLG, the established Boltzmann-transport result is
\begin{equation}
\frac{1}{\tau_{\rm AP}^{\rm MLG}}
=\frac{D_{\rm AP}^{2}k_BT}
{4\rho v_l^{2}\hbar^{3}v_F^{2}}
E_k,
\label{eq:MLGAP}
\end{equation}
which shows that
\(\frac{1}{\tau_{\rm AP}^{\rm MLG}}\propto E_kT.
\)
This energy dependence follows from the linear Dirac dispersion
$E_k=\hbar v_Fk$ and the corresponding energy-dependent DOS. The monolayer chiral overlap factor
$u_{\rm MLG}(\theta)$ also enters the transport angular average.
\cite{hwang2008acoustic}

A more general treatment retaining the finite AP energy
and PB was developed by Khatoon \emph{et al.}
\cite{arshia2021inelastic}. Their semi-inelastic treatment contains
separate phonon absorption and emission contributions and approaches
the quasi-elastic result in the appropriate high-temperature limit.

For low-energy BLG, the corresponding quasi-elastic result obtained
from the present formalism is
\begin{equation}
\frac{1}{\tau_{\rm AP}^{\rm BLG}}
=
\frac{D_{\rm AP}^{2}m^*k_BT}
{4\rho v_l^{2}\hbar^{3}},
\label{eq:BLGAP}
\end{equation}
so that
\(
\frac{1}{\tau_{\rm AP}^{\rm BLG}}\propto T.
\)
Unlike MLG, the BLG rate is independent of carrier energy within the
low-energy parabolic two-band approximation. This behavior follows
from the approximately constant BLG DOS.
\cite{das2010theory,mccann2013electronic}

For a conventional parabolic 2DEG with no chiral overlap factor, the
corresponding quasi-elastic result is
\begin{equation}
\frac{1}{\tau_{\rm AP}^{\rm 2DEG}}
=
\frac{D_{\rm AP}^{2}m_bk_BT}
{2\rho v_l^{2}\hbar^{3}},
\label{eq:2DEGAP}
\end{equation}
and therefore
\(\frac{1}{\tau_{\rm AP}^{\rm 2DEG}}\propto T.
\)
\cite{ando1982electronic}

Thus, the quasi-elastic AP relaxation rates exhibit the
scaling
\begin{equation}
\frac{1}{\tau_{\rm AP}^{\rm MLG}}\propto E_kT,
\qquad
\frac{1}{\tau_{\rm AP}^{\rm BLG}}\propto T,
\qquad
\frac{1}{\tau_{\rm AP}^{\rm 2DEG}}\propto T.
\end{equation}

The identical temperature scaling of BLG and the conventional 2DEG
arises from the equipartition phonon population, while their
parabolic dispersions and approximately energy-independent DOS lead to an energy-independent relaxation rate within the
quasi-elastic approximation. However, their absolute scattering rates
differ because BLG possesses a nontrivial chiral overlap factor,
$u_{\rm BLG}(\theta_{kk'})$, whereas the conventional 2DEG has
$u_{\rm 2DEG}=1$. For the same deformation-potential coupling and
other parameters, the BLG chiral factor reduces the transport angular
average by a factor of two, leading to
\(\frac{1}{\tau_{\rm AP}^{\rm BLG}}
=
\frac{1}{2}
\frac{1}{\tau_{\rm AP}^{\rm 2DEG}}.
\)

The comparison therefore highlights three distinct regimes: MLG
exhibits an energy- and temperature-dependent AP rate,
whereas both BLG and a conventional 2DEG exhibit an energy-independent
rate proportional to temperature within the quasi-elastic
approximation. The difference between BLG and the conventional 2DEG
then arises from the BLG pseudospin chirality, which modifies the
transport angular weighting of AP scattering. 

\subsection{Optical Phonon Scattering}

The BLG OP transport relaxation rate was evaluated both
with and without PB, using the analytical expressions
given in Eqs.~(\ref{eq:OPSS_PB}) and (\ref{eq:OPSS_noPB}), respectively. The analytical PB-inclusive result has been compared with the corresponding direct numerical evaluation and the two results coincide within the numerical accuracy over the
investigated energy range.

 The calculated BLG OP scattering rate, $OP_b$, is shown
in Fig.~\ref{fig:PMBLG}(a), while the corresponding comparison between
MLG and BLG is shown by the $OP_m$ and $OP_b$ curves in
Fig.~\ref{fig:PMBLG}(b). A distinct change in the scattering rate occurs at approximately
\(3.77\times10^{-13}\,\mathrm{erg}\), corresponding to the optical
phonon energy used in the calculation. This feature separates the
regions in which phonon absorption and emission contribute differently
to the scattering rate. At very low carrier energies,
approximately \(0\)--\(2\times10^{-14}\,\mathrm{erg}\), the numerical
scattering points are comparatively sparse, reflecting the restricted
phase space associated with the finite OP energy. In particular, OP emission is kinematically allowed only
when \(E_k\geq\hbar\omega_o\), whereas absorption is allowed at all
carrier energies in the idealized dispersion used here. The resulting energy dependence therefore
differs qualitatively from that of AP scattering, for
which the characteristic phonon energy is much smaller.

 The influence of PB is examined in Fig.~\ref{fig:PMBLG}(a), where the OP relaxation
rates calculated with and without PB are compared. In the Figure solid curves
represent the results without PB, whereas dashed curves
include the PB factor. The rate obtained
without PB is comparatively insensitive to carrier energy except for
the change associated with the phonon-emission threshold. Inclusion of
PB introduces additional energy dependence through the occupation of
the final electronic states. The difference between the two results is
most apparent in the low-energy regime, where the electronic
occupation factors have a stronger influence on the available final
states. Thus, PB provides an important correction to the OP
relaxation rate and should be retained when an accurate inelastic
transport description is required.

The calculated OP relaxation rates of BLG and MLG are
compared in Fig.~\ref{fig:PMBLG}(b). Under the parameters adopted in
the present calculation, the BLG relaxation rate is lower than the
corresponding MLG rate over the investigated energy range. This difference should not be interpreted solely as a difference in
electron--phonon coupling strength, since the scattering rates also
depend on the band dispersion, DOS, chiral overlap factor,
and the specific electron--phonon coupling parameters used for each
system.\cite{castro2009electronic,mccann2013electronic} The comparison nevertheless demonstrates that OP scattering has a different energy dependence and magnitude in
BLG and MLG.

This behavior should also be distinguished from the AP
case. The relative magnitude of the BLG and MLG rates depends on the
phonon branch and on the corresponding electron--phonon coupling
matrix element; therefore, the ordering of the acoustic- and
optical-phonon scattering rates need not be the same.

For comparison, consider first OP scattering in a
conventional 2DEG with parabolic dispersion and an approximately
energy-independent DOS. For a dispersionless OP and a momentum-independent electron--phonon matrix element, the
transport relaxation rate including PB can be expressed as

\begin{equation}
\begin{aligned}
\frac{1}{\tau_{\rm OP}^{\rm 2DEG}}
=
\Gamma_o
\Bigg[
&
N_o
\frac{1-f(E_k+\hbar\omega_o)}
{1-f(E_k)}
\
&+
(N_o+1)
\frac{1-f(E_k-\hbar\omega_o)}
{1-f(E_k)}
\Theta(E_k-\hbar\omega_o)
\Bigg],
\end{aligned}
\label{eq:2DEGOPPB}
\end{equation}
where
 \(\Gamma_o\) denotes the energy-independent scattering prefactor.

In the absence of PB, this expression reduces to
\(\frac{1}{\tau_{\rm OP}^{\rm 2DEG}}
=
\Gamma_o
\left[
N_o+
(N_o+1)
\Theta(E_k-\hbar\omega_o)
\right].
\label{eq:2DEGOPNoPB}
\)
The two terms correspond to OP absorption and emission,
respectively, with emission allowed only when
\(E_{k}\geq\hbar\omega_o\).

The conventional 2DEG and low-energy BLG both possess parabolic
dispersions and approximately energy-independent DOS.
Consequently, their OP scattering rates have similar
phase-space characteristics. The principal distinction is that BLG
contains the chiral overlap factor \(u_{\rm BLG}\), whereas no
corresponding pseudospin factor occurs in a conventional 2DEG.
\cite{ando1982electronic,mccann2013electronic}

In contrast, MLG has the linear Dirac dispersion
and an energy-dependent DOS. An analytical treatment of
OP scattering in MLG including PB was given by Khatoon
\emph{et al.}~\cite{arshia2021inelastic}. Their result may be written, in the notation adopted here, as
\begin{equation}
\begin{aligned}
\frac{1}{\tau_{\rm OP}^{\rm MLG}}
={}&
\frac{V_o^{2}}
{8\rho\omega_o\hbar^{2}v_F^{2}}
\frac{
e^{\frac{E_k}{k_BT}}
\left(
e^{\frac{E_F+E_k}{k_BT}}+1
\right)
}
{
e^{\frac{\hbar\omega_o}{k_BT}}
}
\\[4pt]
&\times
\Bigg[
\frac{
(E_k+\hbar\omega_o)N_o
}
{
e^{\frac{E_F}{k_BT}}
+
e^{\frac{E_k+\hbar\omega_o}{k_BT}}
}
\\[4pt]
&\qquad+
\frac{
(E_k-\hbar\omega_o)(N_o+1)
\Theta(E_k-\hbar\omega_o)
}
{
e^{\frac{E_F+\hbar\omega_o}{k_BT}}
+
e^{\frac{E_k}{k_BT}}
}
\Bigg].
\end{aligned}
\label{eq:MLGOP}
\end{equation}
The first and second terms in Eq.~(\ref{eq:MLGOP}) describe OP absorption and emission, respectively. As in BLG and the
conventional 2DEG, emission is subject to the threshold
\(E_k\geq\hbar\omega_o\). Neglecting PB gives
\begin{equation}
\frac{1}{\tau_{\rm OP}^{\rm MLG}}
=
\frac{V_o^{2}}
{8\rho\omega_o\hbar^{2}v_F^{2}}
\left[
(E_k+\hbar\omega_o)N_o
+
(E_k-\hbar\omega_o)
(N_o+1)
\Theta(E_k-\hbar\omega_o)
\right].
\label{eq:MLGOPNoPB}
\end{equation}

The comparison among 2DEG, BLG, and MLG highlights the combined effects
of band dispersion, DOS, and pseudospin chirality on
OP scattering.\cite{castro2009electronic,
mccann2013electronic,ando1982electronic} The 2DEG and low-energy BLG share
parabolic dispersions and approximately constant DOS,
whereas BLG additionally possesses the chiral overlap factor
\(u_{\rm BLG}(\theta),
\)
MLG, in contrast, has a linear dispersion, an energy-dependent density
of states \(D(E)\propto E\), and the monolayer overlap factor
\(u_{\rm MLG}(\theta_{kk'})\).
Consequently, the OP relaxation rate has different
carrier-energy dependences in the three systems. Nevertheless, the
finite OP energy produces a common emission threshold,
while PB modifies the available final-state phase space in all three systems.

\subsection{Surface Polar Phonon Scattering}

SPP scattering is an intrinsically inelastic carrier-relaxation
mechanism, with distinct absorption and emission channels determined
by the SPP energy $\hbar\omega_s$.
\cite{arshia2021inelastic}
To examine the energy range over which these processes contribute to
carrier relaxation, the BLG SPP scattering rate was evaluated both
with and without PB using the analytical expressions given in
Eqs.~(\ref{eq:SPPSSWF}) and~(\ref{eq:SPPSS}), respectively.

The BLG SPP scattering rate, $SP_b$, is shown in
Fig.~\ref{fig:PMBLG}(a), while the corresponding MLG--BLG comparison
is shown by the $SP_m$ and $SP_b$ curves in
Fig.~\ref{fig:PMBLG}(b). The SPP rate exhibits a distinct change near
the emission threshold. In the present calculation, this occurs at
approximately
\(E_k \simeq \hbar\omega_s,\) corresponding to the SPP energy used in the calculation. Below this
threshold, only phonon absorption contributes, whereas above the
threshold the phonon-emission channel becomes kinematically allowed in
addition to absorption.

The opening of the emission channel produces a pronounced increase in
the total SPP scattering rate. At higher carrier energies, the SPP
rate remains significant, reflecting the continued contribution of
inelastic phonon emission and absorption to carrier relaxation. The
observed threshold behavior follows directly from the energy-
conservation condition contained in the scattering integral.

In the present model, the distance-dependent factors
\(e^{-2qb}
\qquad\text{and}\qquad
e^{-2q(b+l)}\)
account for the attenuation of the substrate-induced SPP electric field
at the positions of the two graphene layers. For the parameter regime
considered here, the layer--substrate separations are sufficiently
small that
\(qb\ll1,
\qquad
q(b+l)\ll1,
\)
and hence
\(
e^{-2qb}\simeq1,
\qquad
e^{-2q(b+l)}\simeq1.
\)
This approximation substantially simplifies the analytical treatment.
The effect of retaining these exponential factors is nevertheless
examined numerically to assess the validity of this approximation.

To further verify the analytical treatment, the fully numerical
scattering integral was compared with the corresponding analytical
expression. The overall agreement between the two results demonstrates
that the analytical evaluation reproduces the numerical transport
integral within the adopted approximations, with the largest
differences occurring near the inelastic threshold.

SPP scattering in MLG has been studied using the Boltzmann transport
formalism, including analytical treatments of the corresponding
inelastic transport integrals in terms of complete elliptic
integrals.\cite{arshia2021inelastic}
The present treatment extends this framework to BLG by incorporating
the low-energy bilayer dispersion, DOS, and chiral overlap factor.

The mathematical structure of the BLG result remains closely related
to that of MLG, with complete elliptic integrals appearing in the
analytical evaluation of the angular dependence. Quantitative
differences, however, arise from the distinct electronic structures.
In particular, low-energy BLG possesses a parabolic dispersion and an
approximately energy-independent DOS, whereas MLG has a linear Dirac
dispersion and an energy-dependent DOS.

\cite{castro2009electronic,mccann2013electronic}
Furthermore, the BLG overlap factor $u_{kk'}$ differs from the MLG
factor $u_{kk'}^{\rm MLG}$. These differences modify both the magnitude
and the carrier-energy dependence of the SPP transport relaxation
rate.

The comparison also highlights the importance of PB. Because SPP
scattering is inelastic, the initial and final electronic states
generally have different energies,
\(E_{k'}=E_k\pm\hbar\omega_s.\)
Consequently, the occupation of the final electronic state can modify
the scattering rate. The present BLG results show that the PB factor
produces a measurable modification of the SPP relaxation rate,
consistent with the role of final-state occupation in inelastic
phonon scattering.

Thus, the present analytical treatment provides a BLG counterpart to
the established MLG SPP framework,\cite{arshia2021inelastic} while
explicitly incorporating the parabolic low-energy BLG band structure
and its associated chiral overlap factor. The resulting formulation
can therefore be used to assess the contribution of substrate-induced
polar phonons to carrier relaxation in supported BLG.

\subsection{Comparative Analysis of Scattering Mechanisms}

The preceding sections have developed analytical transport relaxation
rates for the principal elastic and inelastic scattering mechanisms
considered in BLG, including SR,
AP, OP, SP, NI, and CI scattering. The resulting rates provide a
basis for assessing the relative importance of these mechanisms under
the parameter set adopted in the present calculations.
The scattering mechanisms can be broadly divided into elastic and
inelastic processes. NI, SR, and CI scattering are elastic within the
approximations adopted here, whereas AP scattering is treated in the
quasi-elastic regime and OP and SP scattering involve finite phonon
energies and therefore inelastic carrier transitions. As summarized in
Table~\ref{tab:scattering}, the relative importance of these mechanisms
depends on carrier energy, temperature, carrier density, disorder
strength, dielectric environment, and the relevant electron--phonon or
impurity coupling parameters. Thus, the qualitative classification in
the table does not by itself determine which mechanism dominates for a
particular set of conditions.

\begin{figure}[ht]
\centering
\begin{subfigure}[b]{0.45\textwidth}
\centering
\includegraphics[width=\textwidth]{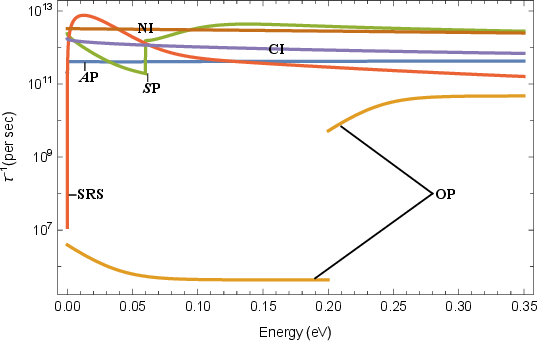}    
\end{subfigure}
\caption{Calculated carrier transport scattering rates, $\tau^{-1}$,
as a function of carrier energy $E_k$ in bilayer graphene for
surface-roughness (SR), acoustic-phonon (AP), surface-polar-phonon
(SPP), neutral-impurity (NI), charged-impurity (CI), and
optical-phonon (OP) scattering mechanisms. The figure compares the
relative magnitudes and energy dependences of the different
scattering channels over the investigated energy range.}
\label{fig:DPBLG}
\end{figure}

Figure~\ref{fig:DPBLG} summarizes the calculated transport scattering
rates as functions of carrier energy and provides a direct comparison
of the relative magnitudes and energy dependences of the six
scattering mechanisms considered here. The figure shows distinct
energy dependences for SRS, AP, SPP, NI, CI, and OP scattering over the
investigated energy range.

A clear hierarchy among the different mechanisms is evident, although
their relative importance is strongly energy dependent. NI scattering
is among the dominant mechanisms over a substantial portion of the
investigated energy range, particularly at intermediate and higher
carrier energies. Its rate remains of the order of
$10^{12}$--$10^{13}~\mathrm{s^{-1}}$ and exhibits only a moderate
variation with carrier energy. This behavior is consistent with the
approximately energy-independent low-energy DOS of BLG and the
finite-range NI transport kernel discussed earlier. The approximately
constant low-energy DOS associated with the parabolic BLG bands is a
well-established feature of its low-energy electronic structure.
\cite{mccann2013electronic,mccann2010electrons,dassarma2010theory}

The AP and CI contributions are also substantial over the investigated
energy range. The AP rate is comparatively weakly dependent on carrier
energy within the quasi-elastic regime, consistent with the
approximately constant low-energy DOS of BLG and the standard
treatment of AP scattering in graphene-based systems.
\cite{hwang2008acoustic,kubakaddi2009interaction,van2020full}
The CI rate decreases progressively with increasing carrier energy in
the present calculation, reflecting the combined effects of carrier
wave vector, dielectric screening, and the chiral transport factor.
CI scattering and its dependence on screening and carrier density have
been extensively studied in MLG and BLG.
\cite{adam2009theory,xiao2010charged,dassarma2010theory}
Thus, although NI scattering is one of the dominant contributions over
a substantial portion of the energy range shown, CI and AP scattering
provide significant parallel momentum-relaxation channels.

For the parameter set adopted here, the SPP contribution is
particularly large in the low-energy region and exhibits a pronounced
change near $E_k\simeq0.06~\mathrm{eV}$, associated with the finite SPP
energy. As discussed in the SPP section, this behavior originates from
the distinction between phonon absorption and the opening of the phonon
emission channel. Above the emission threshold, the total SPP
scattering rate is modified by the additional inelastic phase space.
Consequently, the importance of SPP scattering depends not only on
carrier energy but also on the dielectric environment and the SPP
parameters of the substrate.
\cite{fratini2008substrate,li2010surface,ong2012theory,konar2010effect}

OP scattering is comparatively weak in the lower-energy part of the
investigated range, primarily because the finite OP energy imposes a
threshold for phonon emission. For an OP of energy
$\hbar\omega_{\rm OP}$, emission requires the initial carrier energy to
satisfy approximately
\(E_k\geq\hbar\omega_{\rm OP},
\)
subject to the detailed band structure and momentum-conservation
conditions. Below this threshold, the emission channel is
kinematically inaccessible, and the OP contribution is therefore
strongly suppressed. Once the threshold is reached, the opening of the
inelastic emission channel produces the rapid increase in the OP rate
observed in Fig.~\ref{fig:DPBLG}. Thus, the low-energy weakness of OP
scattering should not be interpreted as an intrinsically weak
electron--optical-phonon coupling in BLG, but rather as a consequence
of the finite phonon energy and the associated inelastic phase space.

The relative importance of AP scattering is further influenced by the
low-energy electronic structure of BLG. Because BLG possesses an
approximately parabolic low-energy dispersion and a nearly constant
DOS, low-energy electronic states provide substantial phase space for
AP scattering. In addition, the multilayer structure introduces
additional acoustic-like phonon branches that can contribute to
carrier scattering. First-principles calculations of electron--phonon
interactions in BLG have shown that low-energy acoustic and
acoustic-like phonon modes can make important contributions to carrier
scattering, particularly at low and moderate carrier energies.
\cite{borysenko2011electron}
Consequently, the larger AP rate than OP rate observed below the OP
threshold in the present calculation results from the combined effects
of the quasi-elastic nature of AP scattering, the available low-energy
phase space, and the finite threshold for OP emission. At sufficiently
high carrier energy, however, OP scattering can become an important
inelastic relaxation channel.
\cite{borysenko2011electron,laitinen2015coupling,arshia2021inelastic}

SRS exhibits a strong energy dependence in the low-energy region. Its
rate rises rapidly from very small values near zero carrier energy and
then decreases progressively with increasing carrier energy. It
therefore provides a non-negligible momentum-relaxation channel over a
considerable portion of the investigated range. This behavior reflects
the dependence of the SRS rate on the Fermi wave vector, the roughness
correlation length, and the associated transport angular kernel.
SRS and interface-induced potential fluctuations are recognized
sources of momentum relaxation in supported low-dimensional systems,
and their importance depends on the microscopic roughness parameters
and device geometry.
\cite{sakaki1987interface,ramayya2008electron,shah2024role}

The results in Fig.~\ref{fig:DPBLG} demonstrate that the dominant
scattering mechanism is not determined by carrier energy alone but by
the combined effects of electronic structure, disorder parameters,
phonon populations, dielectric screening, and inelastic phase space.
In particular, the present calculations show that SPP and SRS can make
large contributions in the low-energy region, while NI becomes one of
the dominant channels over a substantial intermediate- and
higher-energy range. AP and CI provide significant competing
momentum-relaxation channels, whereas OP scattering becomes important
only after its corresponding inelastic emission threshold is
accessible. The relative hierarchy therefore changes with carrier
energy even for the fixed parameter set considered here.

\section{Conclusion}

In this paper, the carrier-scattering characteristics of bilayer graphene (BLG)
have been investigated using complementary analytical and numerical approaches.
The principal scattering mechanisms considered are neutral-impurity (NI),
surface-roughness (SR), charged-impurity (CI), acoustic-phonon (AP),
optical-phonon (OP), and surface-polar-phonon (SPP) scattering. Their
energy-dependent transport relaxation rates were derived within the
Boltzmann transport framework and evaluated numerically over a broad range
of carrier energies. The analysis shows that the relative importance of
these mechanisms is strongly dependent on carrier energy, temperature,
screening, and phonon occupation, and therefore no single scattering
mechanism universally dominates carrier relaxation in BLG.

The NI and SR mechanisms are found to be particularly important in the
low-energy regime. NI scattering exhibits a characteristic energy
dependence associated with the BLG density of states and chiral overlap,
while SR scattering is strongly influenced by the finite correlation length
of the graphene--dielectric interface and by dielectric screening.
Screening substantially modifies the SR relaxation rate and must therefore
be included when assessing interface-limited transport in BLG. The CI
analysis further shows that charged-impurity scattering is strongly
dependent on impurity density and screening, with its low-energy behaviour
reflecting the electronic structure of BLG.

The phonon-mediated mechanisms exhibit distinct energy and temperature
dependences. AP scattering is governed by the available phonon phase space
and the energy dependence of the BLG density of states, while OP scattering
is strongly constrained by the optical-phonon emission threshold and by
Pauli blocking. Consequently, OP scattering becomes increasingly important
once the carrier energy exceeds the relevant phonon energy. SPP scattering
provides an additional inelastic relaxation channel in BLG on polar
dielectric substrates. Its contribution is particularly important in the
low-energy regime considered here and is strongly affected by substrate
properties, phonon energy, carrier energy, and temperature. The comparison
of the analytical and numerical SPP results also demonstrates the
importance of retaining the full energy-dependent BLG density of states
and chiral-overlap factors when evaluating the relaxation rate.

A comparative analysis with monolayer graphene (MLG) and a conventional
two-dimensional electron gas (2DEG) further demonstrates that the
scattering behaviour in BLG cannot be described solely by generic
two-dimensional transport models. The approximately parabolic low-energy
dispersion of BLG gives rise to an approximately constant density of states,
while its chiral overlap factor introduces an additional characteristic
dependence into the scattering rates. These features lead to energy
dependences that differ from those of both MLG and a conventional parabolic
2DEG. The comparison therefore highlights the importance of incorporating
the actual electronic structure of BLG when modelling carrier relaxation.

Overall, the results establish that carrier transport in BLG is governed by
a competition among elastic and inelastic scattering mechanisms whose
relative contributions vary with the physical and material parameters of
the system. The analytical expressions and numerical results presented
here provide a unified framework for identifying the dominant relaxation
channels under different transport conditions and for assessing the roles
of disorder, dielectric screening, and phonon-mediated processes in BLG-based
electronic and high-frequency devices.

\printbibliography
\end{document}